\documentclass[fceqn, useAMS, usenatbib, onecolumn]{mn2e}
\usepackage{epsfig}
\usepackage{amsmath}
\usepackage{multirow}
\usepackage{hyperref}
\def\be{\begin{equation}}
\def\ee{\end{equation}}
\def\ba#1\ea{\begin{align}#1\end{align}}

\def\go{\mathrel{\raise.3ex\hbox{$>$}\mkern-14mu
             \lower0.6ex\hbox{$\sim$}}}
\def\lo{\mathrel{\raise.3ex\hbox{$<$}\mkern-14mu
             \lower0.6ex\hbox{$\sim$}}}

\def\apj{Astrophys. J.}
\def\apjl{Astrophys. J. Lett.}

\def\aap{Astron. Astrophys. }

\def\physrep{Phys. Rep. }
\def\mnras{Mon. Not. Roy. Astron. Soc. }
\def\pasj{Pub. Astron. Soc. Jap.}

\def\prd{Phys. Rev. D.}

\DeclareMathOperator{\F}{F}
\DeclareMathOperator{\E}{E}
\DeclareMathOperator{\am}{am}
\DeclareMathOperator{\sgn}{sgn}
\DeclareMathOperator{\sn}{sn}
\DeclareMathOperator{\cn}{cn}

\def\tomega{\tilde\omega}
\def\tomega{{\tilde{\omega}}}

\begin{document}
\title[Iron Line Variability]
{Iron Line Variability of Discoseismic Corrugation Modes}
\author[D. Tsang and I. Butsky]
{David Tsang$^{1,2}$\thanks{Email:
dtsang@physics.mcgill.ca} 
and Iryna Butsky$^{1}$\footnotemark[1] \\ 
$^1$TAPIR, California Institute of Technology, M.C. 350-17, 1200 E. California Blvd., Pasadena, CA, 91125USA  \\
$^2$Physics Department, McGill University, 3600 rue University, Montreal, QC H3A 2T8, Canada}

\label{firstpage}
\maketitle

\begin{abstract}
Using a fast semi-analytic raytracing code, we study the variability of relativistically broadened Fe-K$\alpha$ lines due to discoseismic oscillations concentrated in the inner-most regions of accretion discs around black holes. The corrugation mode, or c-mode, is of particular interest as its natural frequency corresponds well to the $\sim 0.1-15$Hz range observed for low-frequency quasi-periodic oscillations (LFQPOs) for lower spins. Comparison of the oscillation phase dependent variability and QPO-phase stacked Fe-K$\alpha$ line observations will allow such discoseismic models to be confirmed or ruled out as a source of particular LFQPOs. The spectral range and frequency of the variability of the Fe-K$\alpha$ line due to corrugation modes can also potentially be used to constrain the black hole spin if observed with sufficient temporal and spectral resolution. 
\end{abstract}
\begin{keywords}
accretion, accretion discs -- hydrodynamics -- waves -- black hole physics -- X-rays:binaries -- line:profiles
\end{keywords}

\section{Introduction}
Quasi-periodic oscillations (QPOs) have been detected in the rapid variability of X-ray flux from X-ray binary systems and galactic nuclei for decades. The advent of dedicated timing instruments such as NASA's \emph{Rossi X-ray Timing Explorer} (RXTE) \citep{Swank1999} allowed detailed study of both high and low frequency QPOs in galactic X-ray binaries and AGN. For X-ray binary systems, low-frequency QPOs (LFQPOs) range in frequency from $\sim 0.1- 15$ Hz, with high amplitude and coherence ($Q > 10$), but may vary in frequency over a short (minute) time scale. The high-frequency QPOs (HFQPOs) ranging from $\sim40-450$Hz, have stable frequencies, but low coherence ($Q \sim 2-10$). While much progress has been made in the observations of QPO phenomena, the origins of QPOs remain a mystery, with many theoretical models proposed \citep[see][and references therein for review]{Lai2009}, but thus far none have been convincingly confirmed by observation. 

Perhaps the most theoretically appealing model of QPO behaviour is that of discoseismic oscillation first proposed by \citet{Kato1980} \citep[see][for review]{Wagoner1999, Kato2001}. In these models natural oscillation modes of thin discs in general relativistic potentials are excited at the QPO frequencies. These modes are thought to modify the flux of thermal photons from the disc which are in turn Compton up-scattered to X-ray energies by a hot corona, thereby providing variability in the X-ray flux. Discoseismic models for HFQPOs have been studied extensively \citep[e.g.][]{Nowak1991, Nowak1992, Perez1997, Lai2009}.

While there are several other strong candidate models for LFQPOs [e.g. the large scale modulation of accretion flow \citep{Sobczak2000}, torus modes \citep[e.g.][]{Machida2008}, accretion ejection instabilities \citep[e.g][]{Varniere2012}] in this paper we will focus on the observational signatures of particular discoseismic models of the LFQPOs that may be observable using existing and upcoming instruments. Corrugation modes, or c-modes, are vertical oscillation modes trapped between the inner edge of an accretion disc and the inner vertical resonance \citep[see e.g.][]{Tsang2009}. These modes appear as corrugations in the disc, and typically have very low eigenfrequencies matching the Lense-Thirring precision frequency at the outer edge of their trapping region. As a source of strong perturbation of the inner disc structure with eigenfrequencies in appropriate range, they compelling candidates for the source of LFQPOs. 

Relativistic broadening of iron (Fe-K$\alpha$) lines due to gravitational red shifting and Doppler shifting in black hole accretion discs has been used as a probe of the space time near the inner edge of the disc, with detailed observations of X-ray binaries providing strong constraints on their black hole spins \citep[see e.g.][]{Beckwith2004, Fragile2005}. Fe-K$\alpha$ lines result from the fluorescence of iron atoms within the disc after absorption of incident photons from a nearby X-ray source (usually assumed to be either point sources or a Compton up-scattering hot corona). As the iron line acts as a probe of the inner structure of an accretion disc, strong perturbations or tilting of the inner structure should be reflected in variability  of the iron line. If these perturbations of the inner disc structure are responsible for the LFQPOs, then QPO-phase stacked observations of the iron line emission should produce distinctive signatures, if probed with sufficient temporal and spectral resolution, such as with ESA's proposed LOFT mission \citep{Feroci2012}.

Observationally, correlation between the broadened iron line and QPO phase has been seen in both a galactic X-ray binary, GRS 1915 + 105  using RXTE  \citep{Miller2005}, and a Seyfert 1 galaxy, NGC 3783 using \emph{XMM Newton} \citep{Tombesi2007}. Variability of line profiles due to disc perturbations have previously been modelled by \citet{Karas2001} for a toy model of spiral perturbations in an accretion disc, while \citet{Schnittman2006} explored the line variability due to a precessing tilted ring. While these models do provide a strong source of line variability neither provide a compelling hydrodynamic model for disc perturbation. Here we will instead focus on the line variability due to known oscillation modes of the accretion discs. 

In this paper we perform a detailed calculation of the iron line signatures of discoseismic corrugation mode oscillations. The presence or absence of such QPO phase dependent  signatures will provide strong evidence for the viability of corrugation modes as a source of LFQPOs. In section 2, we outline the calculation of the 
corrugation eigenmodes  of the disc following \citet{Silbergleit2001}, (hereafter SWO). In section 3 we describe how we utilize a fast semi-analytic raytracing code we have developed to calculate the disc images and iron line variability as a function of QPO phase. This code is described in greater detail in Appendix A. We discuss the results of our calculations in section 4, and our conclusions in section 5. 

\begin{figure}
\begin{center}
\includegraphics[height=8cm]{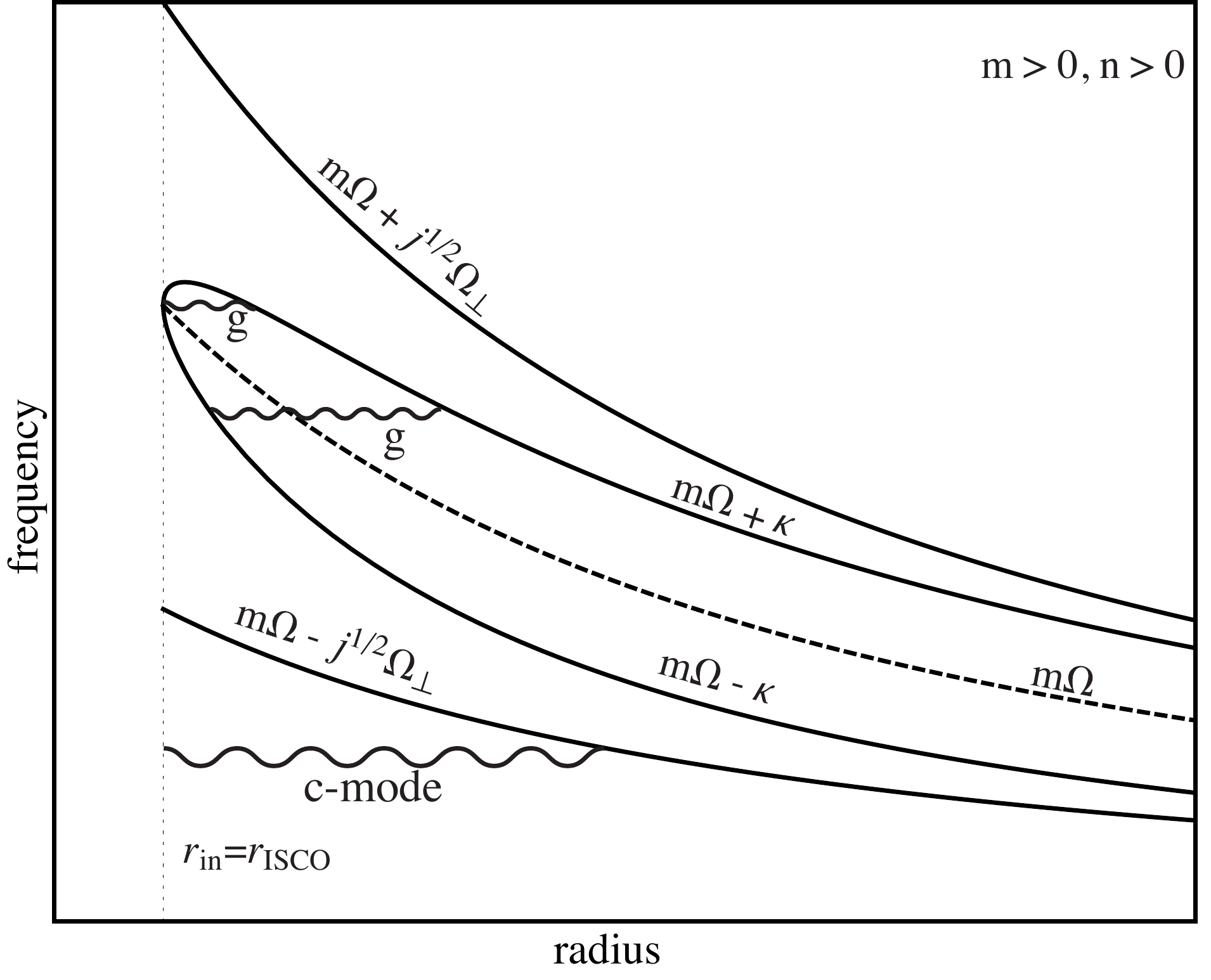}
\end{center}
\caption{Propagation diagram for discoseismic g-modes and c-modes. The c-modes are trapped in the propagation region between the inner disc edge at $\simeq r_{\rm ISCO}$ and the IVR, denoted by the curve $m\Omega - j^{1/2} \Omega_\perp$, while an evanescent region exists between the IVR and the ILR ($m\Omega- \kappa$). The higher-frequency g-modes are trapped in the propagation region between the Lindblad resonances, but are strongly damped if they encounter the corotation resonance. }\label{propdiagram}
\end{figure}

\section{Corrugation Modes}
In units of $G = c = M = 1$ the frequencies of free-particle orbits within a relativistic accretion disc are \citep[e.g.][]{Okazaki1987}
\ba
\Omega &= (r^{3/2} + a)^{-1}\,,\\
\Omega_\perp &= \Omega (1 - 4a/r^{3/2} + 3 a^2/r^2)^{1/2}\,,\\
\kappa &= \Omega (1-6/r  + 8a/r^{3/2} - 3a^2/r^2)^{1/2}\,,
\ea
where $a$ is the dimensionless black hole spin parameter, $\Omega$ is the angular velocity, $\Omega_\perp$ is the vertical epicyclic frequency and $\kappa$ is the radial epicyclic frequency. The inner edge of the disc is located at approximately the innermost stable circular orbit, $r_{\rm ISCO}$, where $\kappa= 0$. 

\begin{figure}
\includegraphics[width=\textwidth]{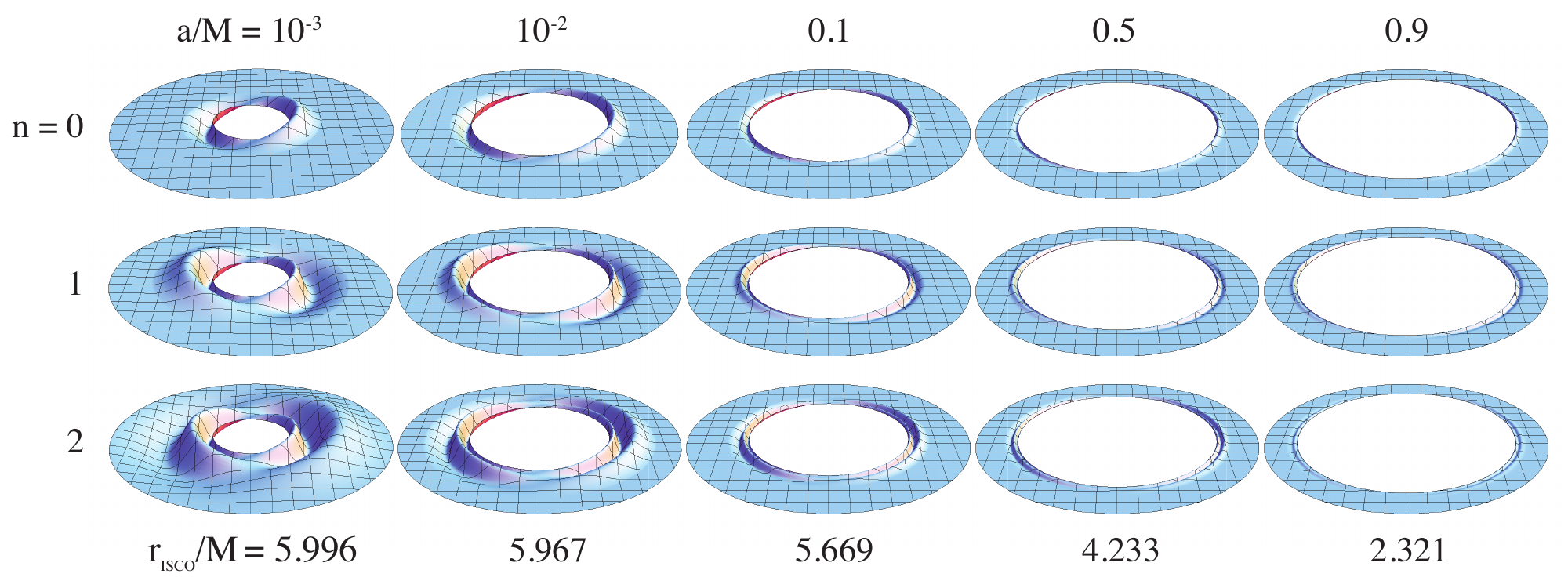}
\caption{The arbitrarily scaled vertical Lagrangian displacements ($\xi^z$) of the fundamental (m=1, j=1) corrugation modes with number of radial nodes $n=0,1,2$ calculated following SWO, for various black hole spin parameters $a/M$.For clarity, the disc images are truncated at $r/M = 22$ for $a/M=10^{-3}$, $r/M=12$ for $a/M = 10^{-2}$, $r/M=10$ for $a = 0.1$, $r/M = 6$ for $a/M=0.5$ and $r/M =3$ for $a/M=0.9$, as the capture region for modes is quite small for higher spin systems. For these systems the disc scale height is set to $H/M=0.01$, adiabatic index $\Gamma=4/3$, with inner boundary phase parameter $\vartheta_{\rm in} = \pi/2$. }\label{cmodes3Dfig}
\end{figure}

Oscillations of frequency $\omega$, azimuthal wave number $m$, and vertical wave number $j$ have critical resonant locations in the disc. These are the Lindblad resonances , where the co-moving perturbation frequency, $\tomega \equiv \omega - m\Omega$, matches the radial epicyclic frequency ($\kappa^2 - \tomega^2 = 0$); the corotation resonance where the pattern speed matches the background fluid speed ($\tomega = 0$); and the vertical resonances, where the co-moving perturbation frequency matches the vertical epicyclic frequency ($j \Omega_\perp^2 - \tomega^2 = 0$).

In the pseudo-Newtonian case, far from the resonances, the approximate WKB dispersion relation is \citep{Okazaki1987}
\be
c_s^2 k^2 \simeq \frac{(\kappa^2 - \tomega^2)(j\Omega_\perp^2 - \tomega^2)}{\tomega^2}
\ee
where $c_s$ is the sound speed, and $k$ is the radial wave number. Corrugation modes, or c-modes, are low frequency modes of accretion discs with vertical structure ($j\neq 0$) that have propagation region between the disc inner edge, and the inner vertical resonance (IVR) (see Figure \ref{propdiagram}). Here we study the fundamental c-modes ($m= j = 1$) which have oscillation frequency equal to the Lense-Thirring precession frequency at the outer edge of their propagation regions, the inner vertical resonances. Higher frequency inertial modes, or g-modes can also be excited in the disc in the region between the Lindblad resonances, where $m\Omega - \kappa < \omega < m\Omega + \kappa$, however, they are quickly damped out by absorption at the corotation resonance unless they exist in the small region where $\kappa$ peaks. Corrugation modes are also damped by interaction with the corotation resonance but at a much smaller rate \citep{Tsang2009}. 

The corrugation modes can be modelled relativistically in the Cowling approximation (no self-gravity) by utilizing the formalism of \citet{Ipser1992}. Here we follow the procedure of SWO in which the relativistic perturbation equations are separated to leading order, using a separation function $\Psi$.

\subsection{Basic Equations}
As in SWO we take background disc to be a relativistic thin accretion disc \citep{Novikov1973} with scale height  $H  \ll r$. We take the background space-time to be Kerr with standard metric components $g^{\mu\nu}$, and utilize Boyer-Linquist coordinates unless otherwise specified.
The background disc fluid four velocity is
$u_o{}^\nu = \beta(t^\nu + \Omega \phi^\nu)$
where
\be
\beta = \frac{r^{3/2} + a}{r^{3/4} ( r^{3/2} - 3r^{1/2} + 2 a)^{1/2}}.
\ee
and $t^\nu$, $r^\nu$, $\phi^\nu$, and $\theta^\nu$ are the Boyer-Lindquist coordinate unit vectors. The sound speed in the disc is given by $c_s = \Gamma^{1/2} H \beta \Omega_\perp$, where $\Gamma > 1$ is the adiabatic index. 

For hydrostatic equilibrium the vertical density and pressure profiles are given by
\ba
\rho &= \rho_o (r) (1-y^2)^{1/(\Gamma - 1)},\\
 p &= p_o(r) (1-y^2)^{\Gamma/(\Gamma -1)},
\ea
where the y-coordinate has been defined such that
\be
y \equiv \frac{z}{H(r)} \sqrt{\frac{\Gamma-1}{2\Gamma}}~.
\ee
For a thin disc the inner region will be dominated by radiation pressure ($\Gamma = 4/3$), with nearly constant scale height $H$ \citep{Novikov1973}.

\begin{table*}
\centering
\begin{minipage}{80mm}
\caption{Properties of the fundamental c-modes for various black hole spins.For these systems the disc scale height is set to $H/M=0.01$, adiabatic index $\Gamma=4/3$, with inner boundary phase parameter $\vartheta_{\rm in} = \pi/2$. \label{modetable}}
\begin{tabular}{| c c c c c c |}
\hline\hline
$a$
& $n$
& $\omega ({\rm rad}~ {\rm s}^{-1}\, M_{10}^{-1})$ 
& $r_{\rm ISCO}/M $
&$ r_{\rm IVR}/M$  
& $\Delta r/M$ \\ \hline\hline
\multirow{3}{*}{$10^{-3}$}
&$0$ 
& $0.0665$
&$5.9967$  
& $8.4880$ 
& $2.4913$  \\
& $1$ 
& $0.0245$
&$5.9967$  
& $11.834$ 
& $5.8380$  \\
& $2$
& $0.00881$
&$5.9967$  
& $16.651$ 
& $10.654$  \\
 \hline
 \multirow{3}{*}{$10^{-2}$}
&$0$ 
& $1.230$
&$5.9673$  
& $6.9086$ 
& $0.9413$  \\
& $1$ 
& $0.853$
&$5.9673$  
& $7.8051$ 
& $1.8378$  \\
& $2$
& $0.615$
&$5.9673$  
& $8.7074$ 
& $2.7401$  \\
\hline
\multirow{3}{*}{$0.1$}
&$0$ 
& $17.8$
&$5.6693$  
& $6.0548$ 
& $0.3855$  \\
& $1$ 
& $15.2$
&$5.6693$  
& $6.3758$ 
& $0.7065$ \\
& $2$
& $13.4$
&$5.6693$  
& $6.6624$ 
& $0.9931$  \\
\hline
\multirow{3}{*}{$0.5$}
&$0$ 
& $195$
&$4.2330$  
& $4.3987$ 
& $0.1657$  \\
& $1$ 
& $179$
&$4.2330$  
& $4.5301$ 
& $0.2971$  \\
& $2$
& $167$
&$4.2330$  
& $4.6428$ 
& $0.4098$  \\
\hline
\multirow{3}{*}{$0.9$}
&$0$ 
& $1460$
&$2.3209$  
& $2.3878$ 
& $0.0669$  \\
& $1$ 
& $1380$
&$2.3209$  
& $2.4394$ 
& $0.1185$  \\
& $2$
& $1320$
&$2.3209$  
& $2.4828$ 
& $0.1619$  \\
 \hline \hline
\end{tabular} 
\end{minipage}
\end{table*}

Using the formalism of \citet{Ipser1992}, following SWO, \citet{Perez1997} and \citet{Nowak1992} and assuming a barotropic disc and perturbation we can write the perturbation equations in term of the enthalpy perturbation $\delta h = \delta p/\rho$ where $p$ is the pressure and $\rho$ is the fluid density. Here the form of the perturbation is given by $\delta \propto \exp[i(m\phi - i\omega t)]$. Defining the Doppler shifted perturbation frequency to be $\tomega \equiv \omega - m\Omega$, we follow SWO  and use the convenient variable
\be 
\delta V \equiv \frac{\delta h}{\beta \tomega} = V_r(r) V_y(r, y),
\ee
where $V_y$ varies only slowly with $r$. 
The coupled ordinary differential equations for the perturbation are then
\ba
(1-y^2) \frac{d^2V_y}{dy^2} - \frac{2y}{\Gamma - 1}\frac{dV_y}{dy} + \frac{2\tomega_*^2}{\Gamma -1} \left[1 + \frac{\Psi - \tomega_*^2}{\tomega_*^2} (1-y)^2\right] V_y &= 0,\label{Vyeqn}\\
\frac{d^2 V_r}{dr^2} - \frac{1}{\tomega^2 - \kappa^2} \frac{d}{dr}(\tomega^2 - \kappa^2) \frac{dV_r}{dr} + \alpha^2 (\tomega^2 - \kappa^2) \left( 1 - \frac{\Psi}{\tomega_*^2}\right)V_r &= 0, \label{Vreqn}
\ea
where $\tomega_* \equiv \tomega/\Omega_\perp$,
$\alpha^2 \equiv \beta^2 g_{rr}/c_s^2$, and $\Psi$ is the separation function which varies slowly in r, with $\Psi\simeq \omega_*^2$ in the propagation region for the c-mode. 

\subsection{WKB Solution}
The WKB solution given by SWO is valid to order 
\be
\epsilon \equiv -\frac{\omega - m(\Omega - \Omega_\perp)}{m \Omega_\perp}.
\ee
For the corrugation modes the solution to the vertical equation \eqref{Vyeqn} is to lowest order
\be
V_y \propto C_j^\lambda (y)
\ee
where $C_j^\lambda$(y) is the Gegenbauer polynomial \citep{AbramowitzandStegun}, with $\lambda = (3-\Gamma)/2(\Gamma-1)$. This gives rise to a selection rule 
$2m^2= j^2(\Gamma-1) + j(3-\Gamma)$, which allows only particular $m$ and $j$ to solutions to exist for a given $\Gamma$. Only the axisymmetric $m=0=j$ mode and the fundamental corrugation mode $m^2=1=j$ exist for all $\Gamma$. For the $m=1$ fundamental corrugation modes we have 
$V_y  \propto y$, with separation function (to linear order) given by
\be
\Psi/\tomega^* \simeq 1 - \epsilon \chi_1
\ee
where $\chi_1 = 3\Gamma - 1$.

%

\begin{figure}
\begin{center}
\includegraphics[width=7.0in]{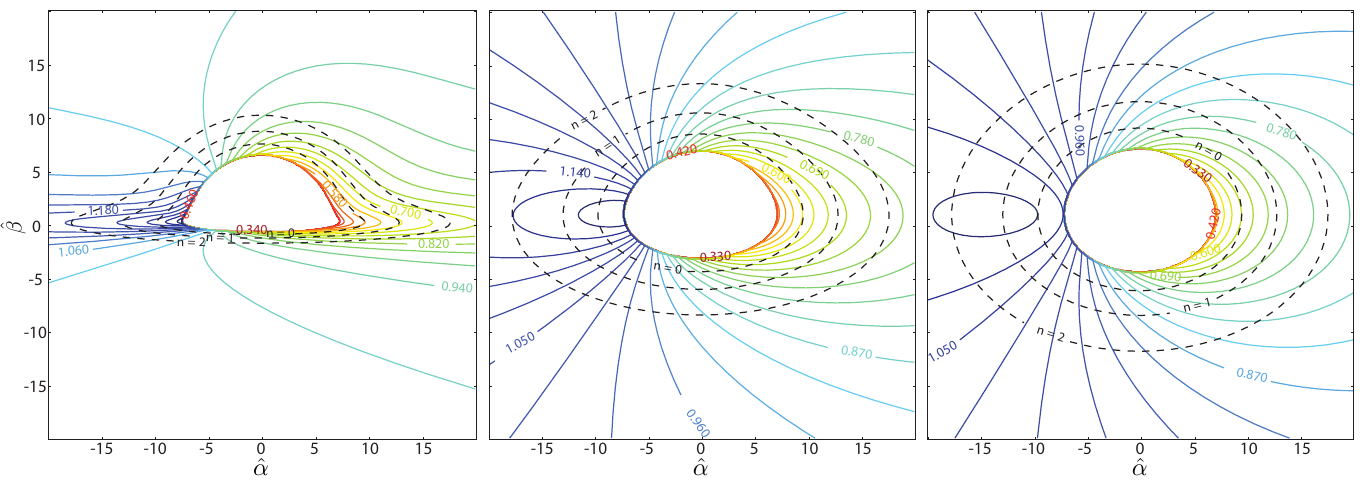} 
\end{center}
\caption{Constant observed frequency ratio ($g = 1/[1 + z]$)  contours for impact parameters $(\hat{\alpha}, \hat{\beta})$ for a disc around a black hole with spin a/M = 0.001 and inclination $\mu_o = 0.1, 0.5, 0.7$ from left to right. The dotted lines represent the inner vertical resonance, $r_{\rm IVR}$, for n = 0, 1, 2 perturbations. The variability of the corrugation modes is confined between the $r_{\rm ISCO}$ and $r_{\rm IVR}$, thus the spectral variation is confined to the redshift bins which have contours within the mode propagation region.  \label{contour001}}
\end{figure}

For the radial equation the solution is given in terms of the $\tau$, which is defined such that
\be
\frac{d\tau}{dr} = \tomega^2 - \kappa^2
\ee
with $\tau(r_{\rm ISCO}) = 0$. The WKB solution is then
\be
V_r \propto Q^{-1/4}(\tau) \cos(\Phi(\tau) + \Phi_i)
\ee
where 
\be
Q(\tau) = \frac{\chi_1 \epsilon \alpha^2}{\tomega^2 - \kappa^2},
\ee
$\Phi_i$ is determined by the inner boundary condition,
and
\be
\Phi(\tau) = \int_0^\tau Q^{1/2} (\tau') d\tau'.
\ee

For the inner boundary condition we utilize a parameterization 
\be
\frac{d}{dr}V_r(r_{\rm ISCO}) \cos \vartheta_{\rm in} - V_r(r_{\rm ISCO}) \sin \vartheta_{\rm in} = 0, \label{BC1}
\ee
where $\vartheta_{\rm in}$ generalizes our ignorance of the boundary condition at ISCO. SWO carefully analyze the singularity at ISCO as the sound speed vanishes, $c_s(r) \rightarrow 0$, which occurs when the torque vanishes at the inner-disc edge, as in the standard Novikov-Thorne model \citep{Novikov1973}. However, recent MHD simulations of accretion discs \citep{Noble2010, Penna2010, Penna2012}, show that for realistic discs including magnetic effects, the torque at ISCO is finite, with $c_s \rightarrow 0$ as $h \rightarrow 0$, due to magnetic stresses connecting the ISCO material to the material in the plunging region. This provides a non-zero sound speed, resulting in well behaved non-singular behaviour at the inner boundary.  In this work, for simplicity, we arbitrarily select the inner boundary $\vartheta_{\rm in} = \pi/2$ such that 
$V_r(r_{\rm ISCO}) = 0$.

With this parameterization we find
\be
\tan \Phi_i= \frac{d \ln Q/dr - \tan(\vartheta_{\rm in})}{Q^{1/2}(\tomega^2 - \kappa^2)}.
\ee
Using asymptotic matching \citep[see e.g.][]{Tsang2008} across the IVR, and taking $n$ to be the number of radial nodes in the trapping region, we can then write the WKB eigenvalue condition,
\be
\Phi_i  +  \int_0^{\tau_{\rm IVR}} Q^{1/2} (\tau') d\tau' = \pi(n - 1/4),
\ee
which can be solved for the approximate mode frequency. 

Using this WKB estimate for the mode frequency as an initial guess, along with the lowest order vertical solution and separation function we can then solve the radial differential equation \eqref{Vreqn} numerically
for the eigenfrequency and eigenmodes using standard shooting methods \citep[e.g.][]{Press1992}. With the boundary conditions given by \eqref{BC1}, and
\be
\bigg(\frac{dV_r}{dr} + \sqrt{-Q}(\tomega^2 - \kappa^2) V_r \bigg)_{r = r_{\rm out}}=0,
\ee
an evanescent decay at some point in the evanescent region, $r_{IVR} < r_{\rm out} < r_{ILR}$, we find the eigenmodes shown in Figure \ref{cmodes3Dfig} and Table \ref{modetable}, for a disc scale height $H/M=0.01$ with different values of black hole spin $a$, and number of radial nodes $n$.

\subsection{Lagrangian Displacements}

The Lagrangian displacements \citep{Nowak1992, Perez1997} are related to the perturbations as
\ba
\xi^r &\simeq -\frac{\tomega g^{rr}}{\beta(\tomega^2 - \kappa^2)} \frac{\partial}{\partial r}\delta V,\\
\xi^z &\simeq -\frac{\tomega}{\beta(\tomega^2 - N_z^2)} \left( \frac{\partial}{\partial z} \delta V + \rho A_z \delta V\right),\\
\xi^\phi &\simeq i\frac{u^t u_t}{\tomega} \left( \frac{\partial \Omega}{\partial r} + \frac{r \nu^z}{\beta^2 \Delta^*} \right)\xi^r,
\ea
where $\Delta^* \equiv r^2 - 2r + a^2$, $\nu^z$ is the vorticity, $N_z$ is the vertical Br\"{u}nt-Vasala frequency, and  $A_z = \beta N_z^2/\partial_z p$. For the barotropic case considered here $A_z = N_z = 0$ and we have the vertical Lagrangian displacement
\be
\xi^z \simeq -\frac{1}{\beta \tomega}\frac{\partial}{\partial z}\delta V.
\ee

\subsection{Tilted Discs}
To provide a comparison we also examine the line variability of a simplified tilted disc model, where we assume that the thin inner disc is tilted by a small amount and precessing as a solid body with the QPO oscillation frequency $\omega$. 
Here we model a precessing slightly tilted disc simply with a vertical lagrangian perturbation given by
\be
\xi^z = A r \cos(\phi - \omega t),
\ee
where $\omega$ is the precession frequency and we set the tangent of the tilt angle $A = 0.01$. We arbitrarily take $\omega$ to be the same frequency as for the $n=0$ c-mode for each spin $a$. While this toy model is not a true hydrodynamical model of the disc, similar disc tilts have been seen in simulations by \citet{Fragile2007}, with the disc globally precessing rather than warping due to  the Bardeen-Petterson effect \citep{Bardeen1975}.

\begin{figure}
\begin{center}
\includegraphics[width=7.0in]{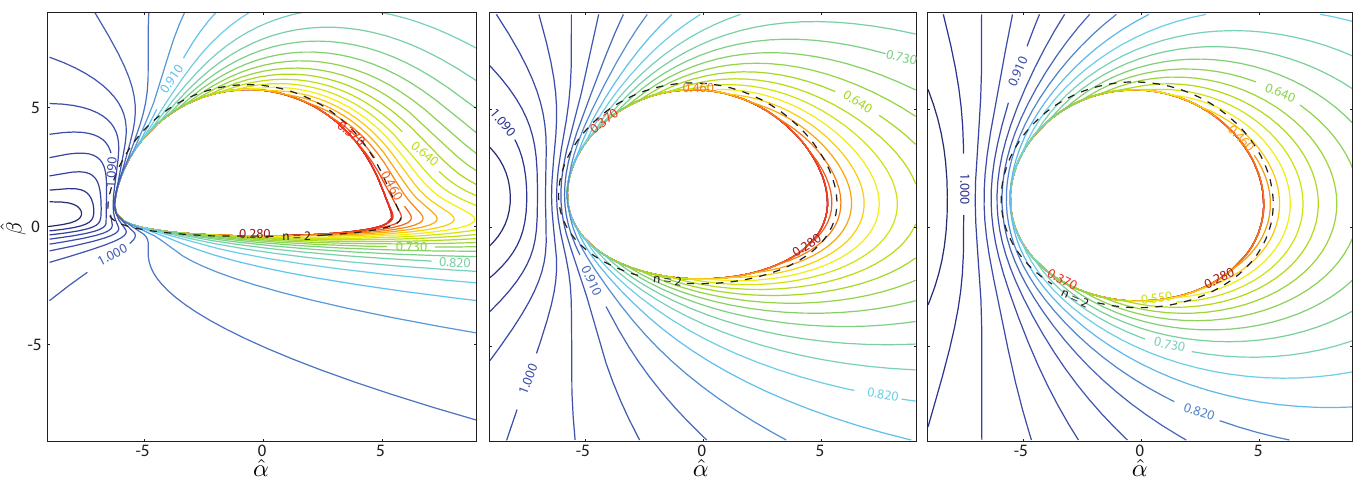} 
\end{center}
\caption{Constant observed frequency ratio ($g = 1/[1 + z]$) contours for impact parameters $(\hat{\alpha}, \hat{\beta})$ for a disc around a black hole with spin a/M = 0.5 and inclination $\mu_o = 0.1, 0.5, 0.7$ from left to right. The dotted lines represent the the inner vertical resonance, $r_{\rm IVR}$, for n = 2 perturbations.  In this region, close to $r_{\rm ISCO}$, the gravitational redshift dominates the orbital Doppler boosting. Since the corrugation-mode propagation region is small for high spin black holes, the redshift range with spectral variability will be correspondingly small.  \label{contour5}}
\end{figure}

\section{Calculating Observables}

To calculate the effect of such perturbations on the disc spectrum and image, we developed a new fast semi-analytic raytracing method\footnote{This code is available for download at \url{http://www.tapir.caltech.edu/~dtsang/qpotrace.html}.} \citep[based on work from][]{TsangThesis}, which we outline in detail in Appendix \ref{CodeAppendix}. Solving the geodesic equations utilizing the ``Mino parameter'' \citep{Drasco2004}, we can find the position four-vectors $x^{\nu}$ and photon four-momenta $k_{\nu}$ at the surface of the the accretion disc. Here the time component of the position vector $x^{t} = \Delta t$ is the difference in the coordinate time $t$ from a particular datum value (see Appendix \ref{TeqAppendix} for more detail on avoiding the divergent terms) allowing us to assess the relative coordinate time delay observed between different photon paths. In this work we ignore secondary and higher order images, as these contribute relatively little to the overall flux, however this is relatively simple to include in our raytracing methods. 

Along each ray the quantity $I_\nu/\nu^3$ is Lorentz invariant, where $I_\nu$ is the specific intensity, and $\nu$ is the photon frequency. The total redshift of the photon is defined as $1 + z \equiv 1/g$, where $g \equiv \nu_{\rm obs}/\nu_{\rm em}$ is the observed frequency ratio (not to be confused with the metric components $g^{\mu\nu}$), hence we have $I_{\nu_{\rm obs}} = g^3 I_{\nu_{\rm em}}$ along a particular ray.

Covariantly the observed frequency ratio, can be given by:
\be
g = \frac{k^{\rm obs}{}_\mu u_{\rm obs}{}^\mu}{k^{\rm em}{}_\nu u_{\rm em}{}^\nu}
\ee
for a distant stationary observer with four velocity $u_{\rm obs}{}^\nu$, where $u_{\rm em}{}^\nu$ is the four velocity of the disc material and $k^{\rm em}{}_\nu$ and $k^{\rm obs}{}_\nu$ are the photon four-momenta at the disc and observer respectively.

To calculate images we divide up the solid angle subtended by the black hole disc in the observer's sky into pixels, denoted by $\hat{\alpha}$, the impact parameter perpendicular to the spin axis, and $\hat{\beta}$, the impact parameter parallel to the spin axis \citep[see e.g.][]{Beckwith2004}.

\begin{figure}
\begin{center}$
\begin{array}{cc}
\includegraphics[width=3.4in]{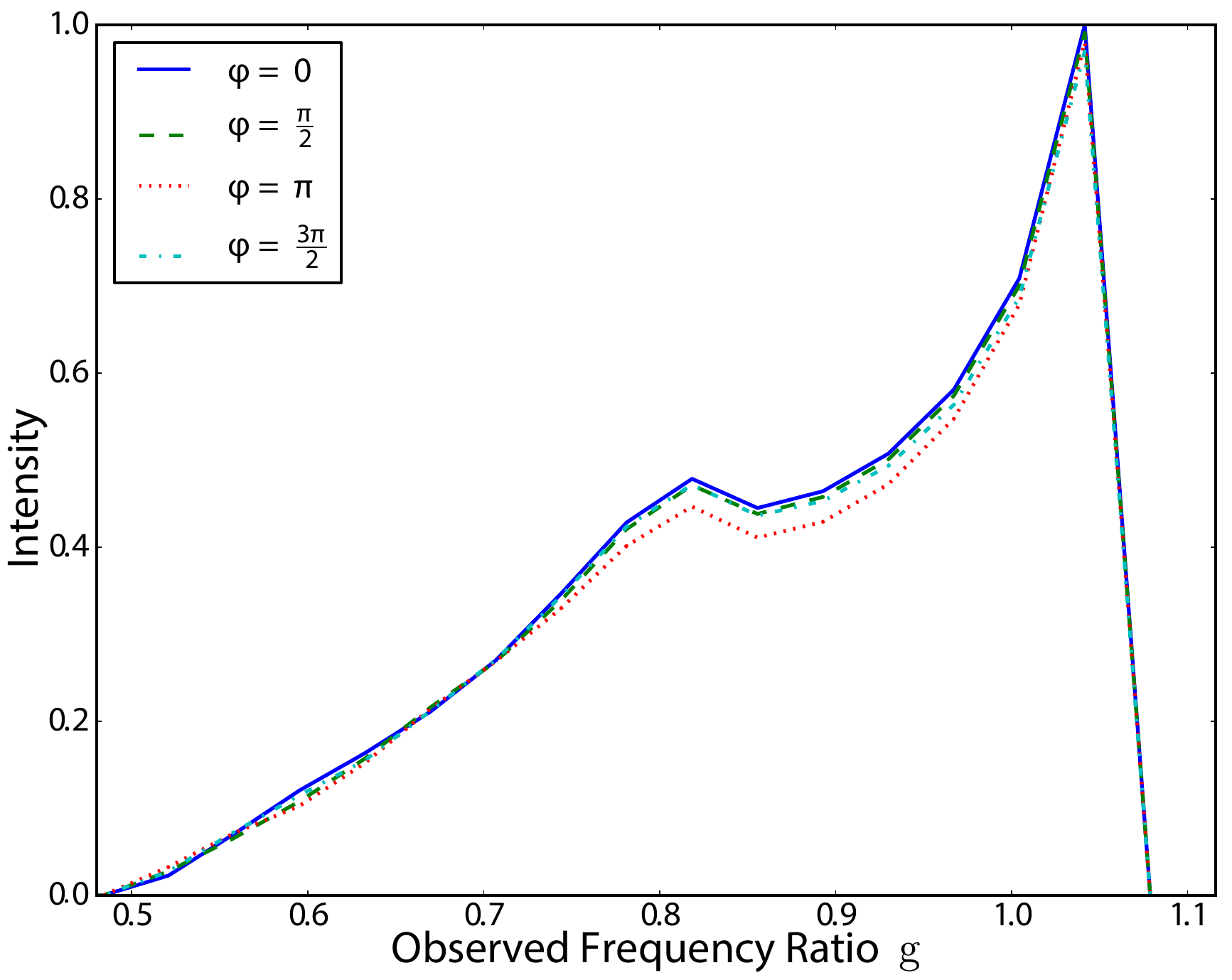} &
\includegraphics[width=3.4in]{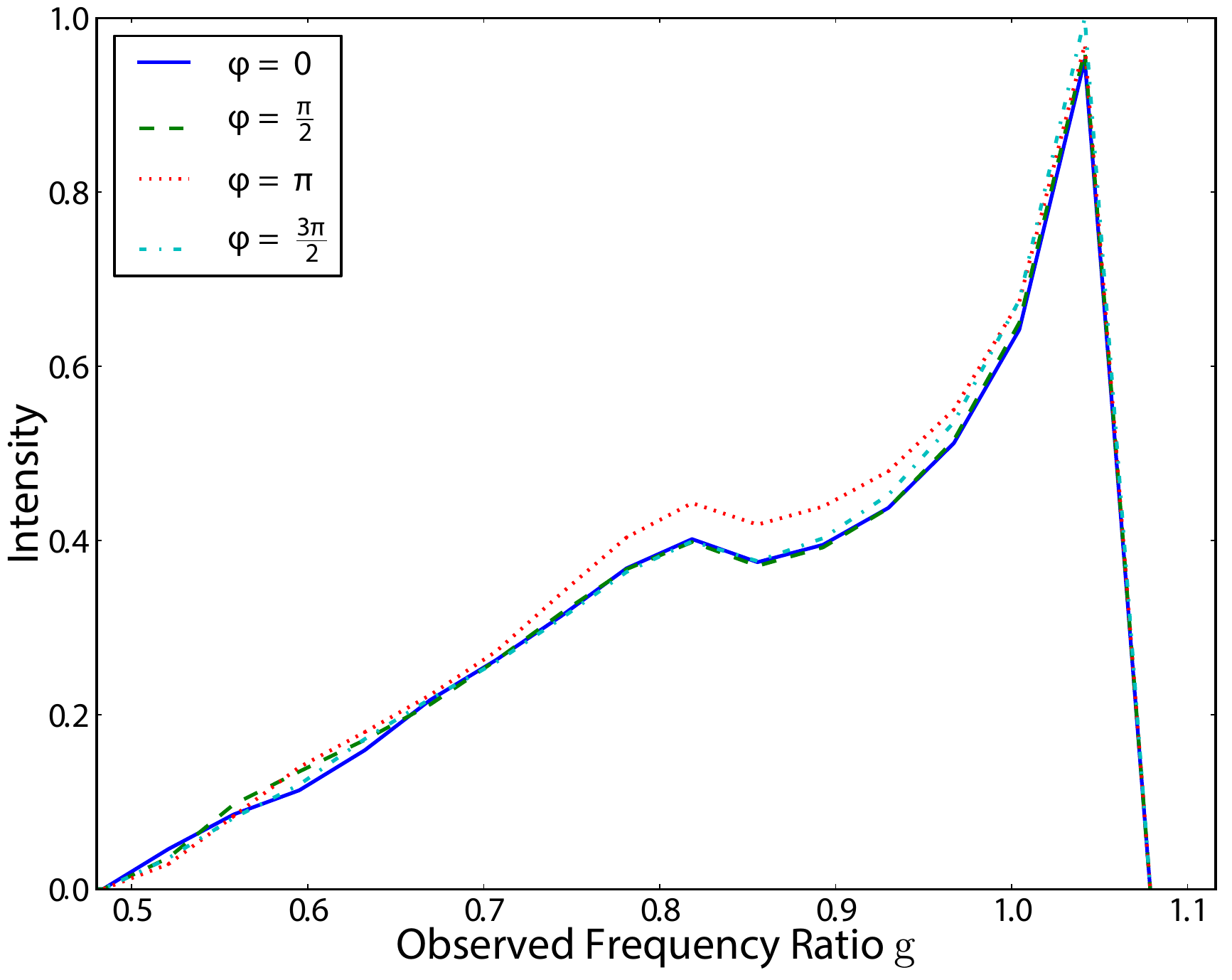}
\end{array}$
\end{center}
\caption{The relative intensity (arbitrary scale) of line emission for system with black hole spin $a = 0.001$, and inclination $\mu_o = 0.7$,  as a function of observed frequency ratio $g \equiv \nu_{\rm obs}/\nu_{\rm em} = 1/(1 + z)$, for limb darkening emissivity (left) and limb brightening emissivity (right).  The broadened lines are shown as a function of  $\phi$, the phase of the applied $n=2$ perturbation with maximum amplitude $\xi_{z, max} \simeq 0.7M$ located at $r \simeq 6.8 M$. \label{specfig}}
\end{figure}

The most important observable to be calculated for each pixel is the observed flux. We have
\ba
F_{\nu_{\rm obs}} &= \int I_{\nu_{\rm obs}} d\Omega\\
&=  \frac{1}{D^2}\int\!\!\!\int g^3 I_{\nu_{\rm em}}d\hat{\alpha} d\hat{\beta} \label{imageeq}
\ea
where $D$ is the distance from the observer to the black hole.
The emissivity of the Fe-K$\alpha$ fluorescence depends on a incident X-ray intensity, number density and ionization state of iron atoms. Since the disc has a vertical thermal and ionization structure, the emission may be subject to an angular dependence. We separate out the angular and radial dependence for the specific emitted intensity
\be
I_{\nu_{\rm em}}(r, \mu_{\rm em}, \nu_{\rm em}) = {\cal R}(r) f(\mu_{\rm em}) \delta(\nu_{\rm em}- \nu_o)  \label{Ieqref}
\ee
where $\mu_{\rm em} = \cos \vartheta_{\rm em}$ is the cosine of the emission angle in the local frame, and $\nu_o$ is the frequency of the fluorescence line in the rest frame. We take the radial dependence of the emissivity to have the form ${\cal R}(r) \propto r^{-q}$, assuming the standard value $q = 3$, which follows the thermal dissipation of the disc \citep{Novikov1973}, however steeper profiles may also be appropriate \citep[see e.g.][]{Svoboda2012}.

\subsection{Angular Emissivity}

\begin{figure}
\begin{center}
\includegraphics[width=6.2in]{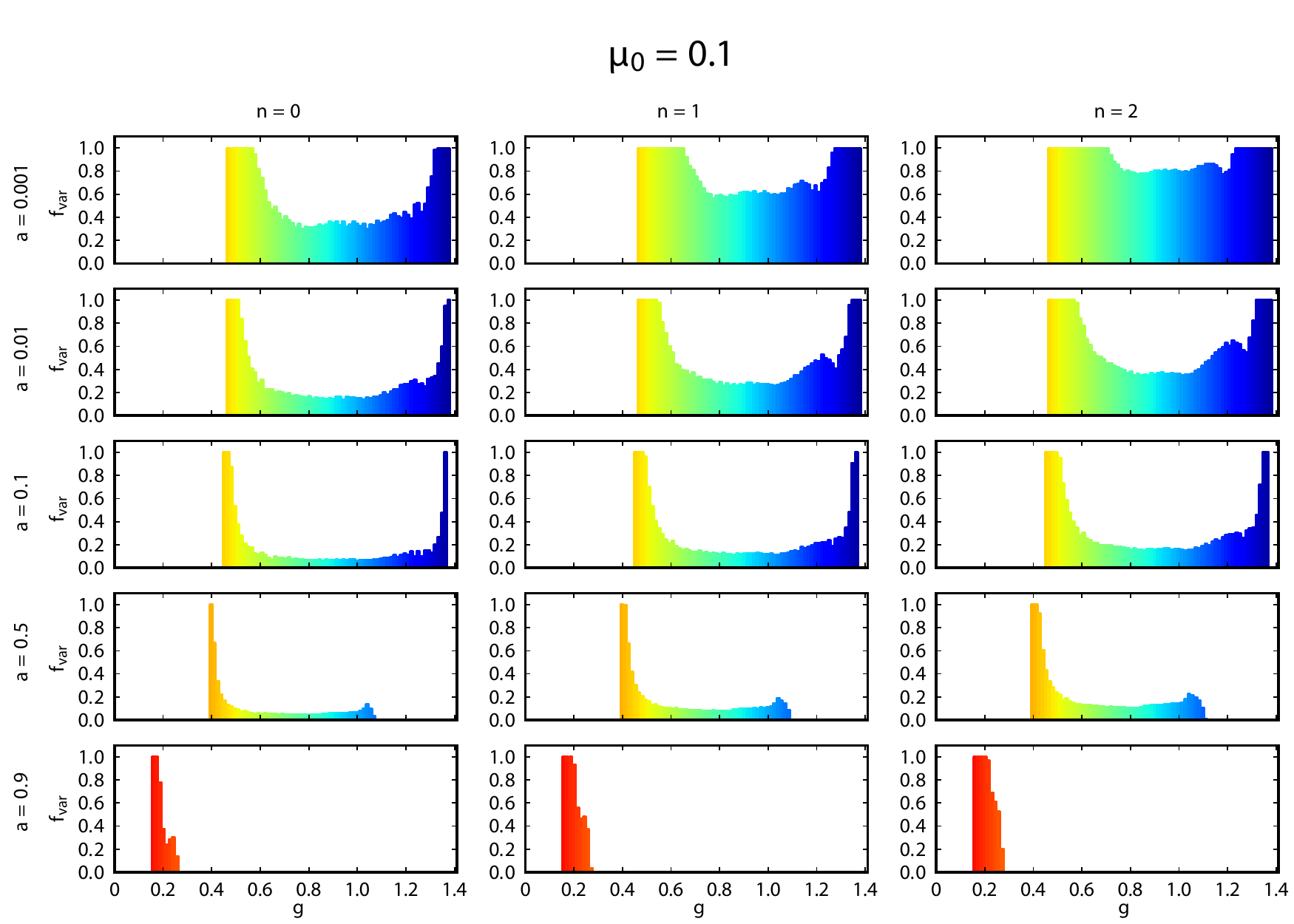}
\end{center}
\caption{The fraction of the intensity ($f_{\rm var}$) that is emitted from the variability region for each observed frequency ratio ($g \equiv 1/[1+z]$) bin. With cosine of the inclination angle $\mu_o = 0.1$, the columns represent perturbations with different numbers of radial nodes (n=0, 1, 2), while the rows represent black hole spin parameters $a = 10^{-3}, 10^{-2}, 10^{-1}, 0.5, 0.9$). For the most redshifted (lowest g) parts of each line profile, the intensity is dominated by emission from the variability region. In this case, we consider angular emissivity to be governed by the limb darkening law, $ f(\mu_{em} = (1 + 2.06\mu_{em})$, with radial emissivity given by ${\cal R}(r) \propto r^{-3}$. For systems with large amplitude perturbations we can expect fractional intensity variation on the order of $f_{\rm var}$ for a given redshift bin.  \label{fvar0.1.fig}}
\end{figure}

The angular emissivity depends on the detailed vertical thermal and ionization structure of the disc atmosphere, which is beyond the scope of this work. For simplicity we will follow \citet{Svoboda2009}, and utilize two different forms of the angular emissivity: a standard limb darkening profile $f(\mu_{\rm em}) = 1 + 2.06\mu_{\rm em}$, \citep{Laor1991}, and a limb brightening profile for a plane parallel atmosphere $f(\mu_{\rm em}) = \ln(1 + \mu_{\rm em}^{-1})$ \citep{Haardt1993}.  We do not consider an isotropic angular emissivity since  for the c-modes which are mostly incompressible perturbations, such an angular profile would result in negligible variability.

The emission angle $\vartheta_{\rm em} = \cos^{-1} \mu_{\rm em}$ is defined as the spatial angle in the local frame between the emitted photon and the normal to the surface of the accretion disc.
The unit normal $\hat{n}$ is  determined using a scalar surface function $F(x^{\alpha})$ and the projected spatial derivative in the frame co-moving with the disc material
\be
n_\mu \equiv \nabla_\mu F(x^{\alpha}) + u_\mu u^\nu \nabla_\nu F(x^\alpha), \qquad \hat{n}_\mu \equiv \frac{n_\mu}{n_\nu n^{\nu}}
\ee
where $u^\nu$ is the background four velocity of the disc material. For the unperturbed case we can use the surface function $F({\bf x}) = r\cos \theta = 0$ which defines the surface of the equatorial plane, giving $\hat{n}^\nu_o = -(1/r) \theta^\nu$, where $\theta^\nu$ is the unit vector in the $\theta$-direction at the equatorial plane. The emission angle is then given as
\be
\mu_{\rm em} = \cos \vartheta_{\rm em} \equiv \frac{\hat{n}_\mu k^{\mu}}{u_{\nu} k^{\nu}}~,
\ee
where all quantities are evaluated at the emission point of the photon at the disc surface.

To extend this to the corrugation mode we define a vertical perturbation of the surface $\xi^z = \xi^z(t, r, \phi)$. The normal vector is now defined using the surface function
\be
F(x^\alpha) = r \cos \theta - \xi^z(t, r, \phi)  = 0
\ee 
Since the velocity of the perturbation is small compared to the background Keplerian flow, we can continue to use the approximation $u^\nu \simeq u_o{}^\nu$. For simplicity we will also approximate the surface displacement $\xi^z$ with the vertical Lagrangian displacement at the disc mid-plane. We also limit our calculations to small perturbations in the limb brightening case such that this approximation does not produce singular values for the angular emissivity.

\begin{figure}
\begin{center}
\includegraphics[width=6.2in]{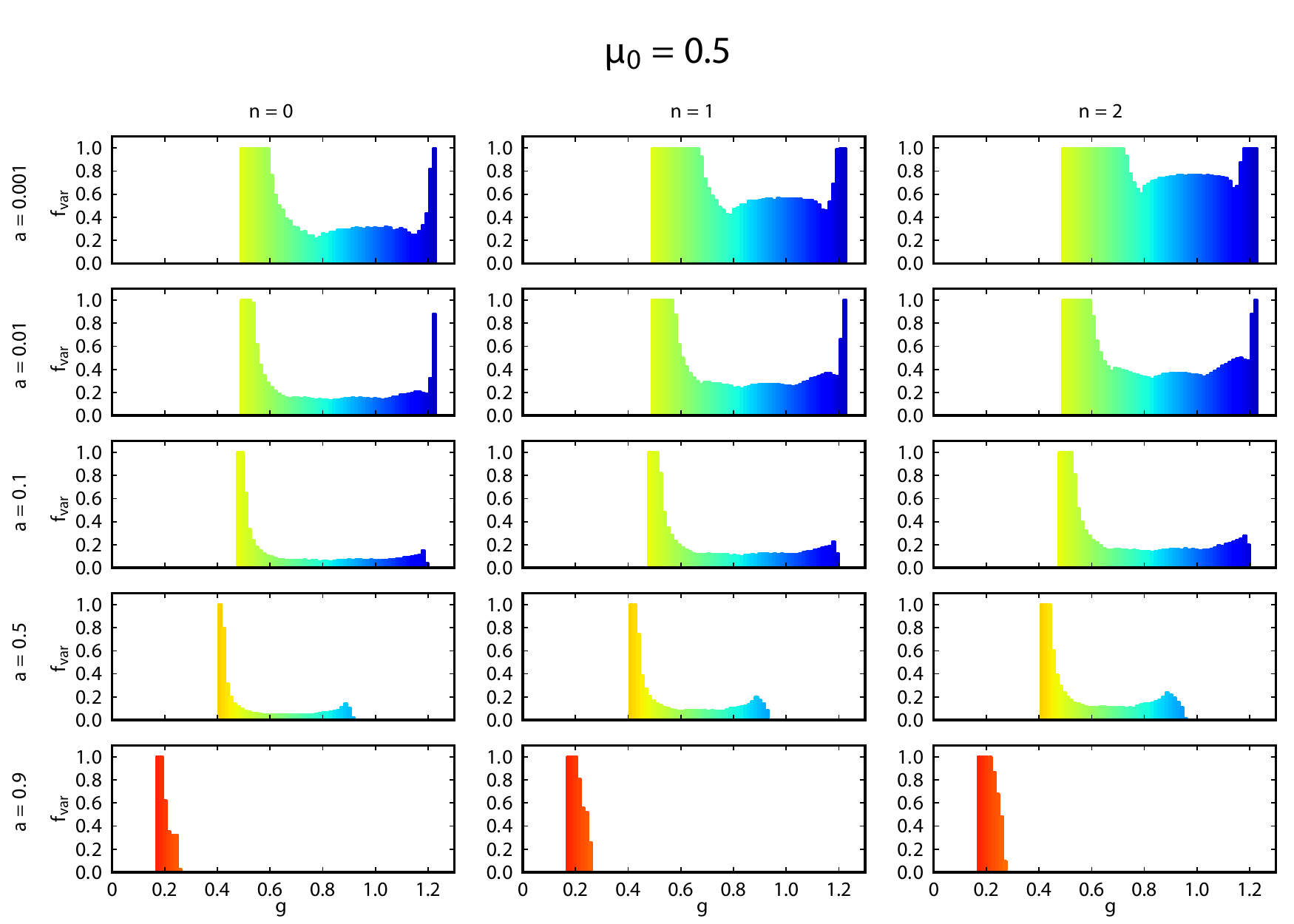}
\end{center}
\caption{The fraction of the intensity ($f_{\rm var}$) that is emitted from the variability region for each observed frequency ratio ($g \equiv 1/[1+z]$) bin. With cosine of the inclination angle $\mu_o = 0.5$, the columns represent perturbations with different numbers of radial nodes (n=0, 1, 2), while the rows represent black hole spin parameters $a = 10^{-3}, 10^{-2}, 10^{-1}, 0.5, 0.9$). For the most redshifted (lowest g) parts of each line profile, the intensity is dominated by emission from the variability region. In this case, we consider angular emissivity to be governed by the limb darkening law, $ f(\mu_{em} = (1 + 2.06\mu_{em})$, with radial emissivity given by ${\cal R}(r) \propto r^{-3}$. For systems with large amplitude perturbations we can expect fractional intensity variation on the order of $f_{\rm var}$ for a given $g$-bin.  \label{fvar0.5.fig}}
\end{figure}

\begin{figure}
\begin{center}
\includegraphics[width=6.2in]{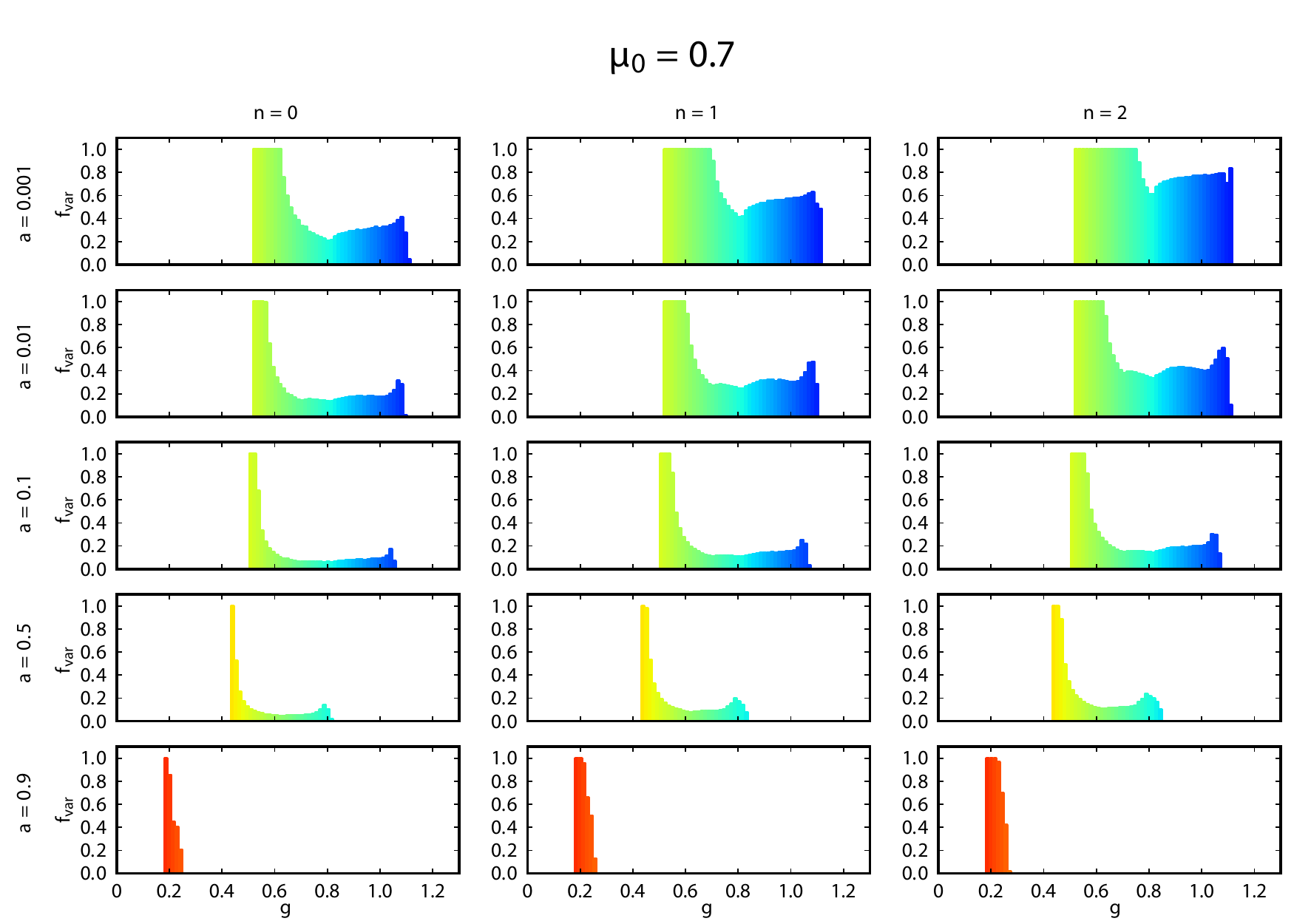}
\end{center}
\caption{The fraction of the intensity ($f_{\rm var}$) that is emitted from the variability region for each observed frequency ratio ($g \equiv 1/[1+z]$) bin. With cosine of the inclination angle $\mu_o = 0.7$, the columns represent perturbations with different numbers of radial nodes (n=0, 1, 2), while the rows represent black hole spin parameters $a = 10^{-3}, 10^{-2}, 10^{-1}, 0.5, 0.9$). For the most redshifted (lowest g) parts of each line profile, the intensity is dominated by emission from the variability region. In this case, we consider angular emissivity to be governed by the limb darkening law, $ f(\mu_{em} = (1 + 2.06\mu_{em})$, with radial emissivity given by ${\cal R}(r) \propto r^{-3}$. For systems with large amplitude perturbations we can expect fractional intensity variation on the order of $f_{\rm var}$ for a given $g$-bin. \label{fvar0.7.fig}}
\end{figure}

\section{Results}


Figures \ref{contour001} and \ref{contour5} show constant redshift ($g = 1/[1+z]$) contours for example systems with black hole spin $a = 0.001$ and $0.5$ respectively. Modulation of the emission at any particular location will modify the observed intensity in that redshift bin. Also shown with dashed lines are the radii corresponding to the inner vertical resonance for particular modes ($n = 0, 1, 2$ for $a = 0.001$ and $n=2$ for $a = 0.5$). The propagation region for each of the corrugation modes is from the ISCO to the IVR. 
In Figures \ref{fvar0.1.fig}-\ref{fvar0.7.fig}, we calculate the fractional intensity $f_{\rm var}$ emitted from the variability region $r < r_{\rm IVR}$ for each spectral bin for the broadened line. All of flux at the reddest parts of the broadened line emerge from the variability region for each case, while for lower spin, the blue wing and a significant portion of the rest of the spectrum should also be variable. For high spin $a = 0.9$, we see that the variability will be confined only to the most redshifted part of the spectrum, as the variability region is very close to the black hole where gravitational redshift strongly dominates the Doppler shift. 

\begin{figure}
\begin{center}$
\begin{array}{c}
\includegraphics[width=6.4in]{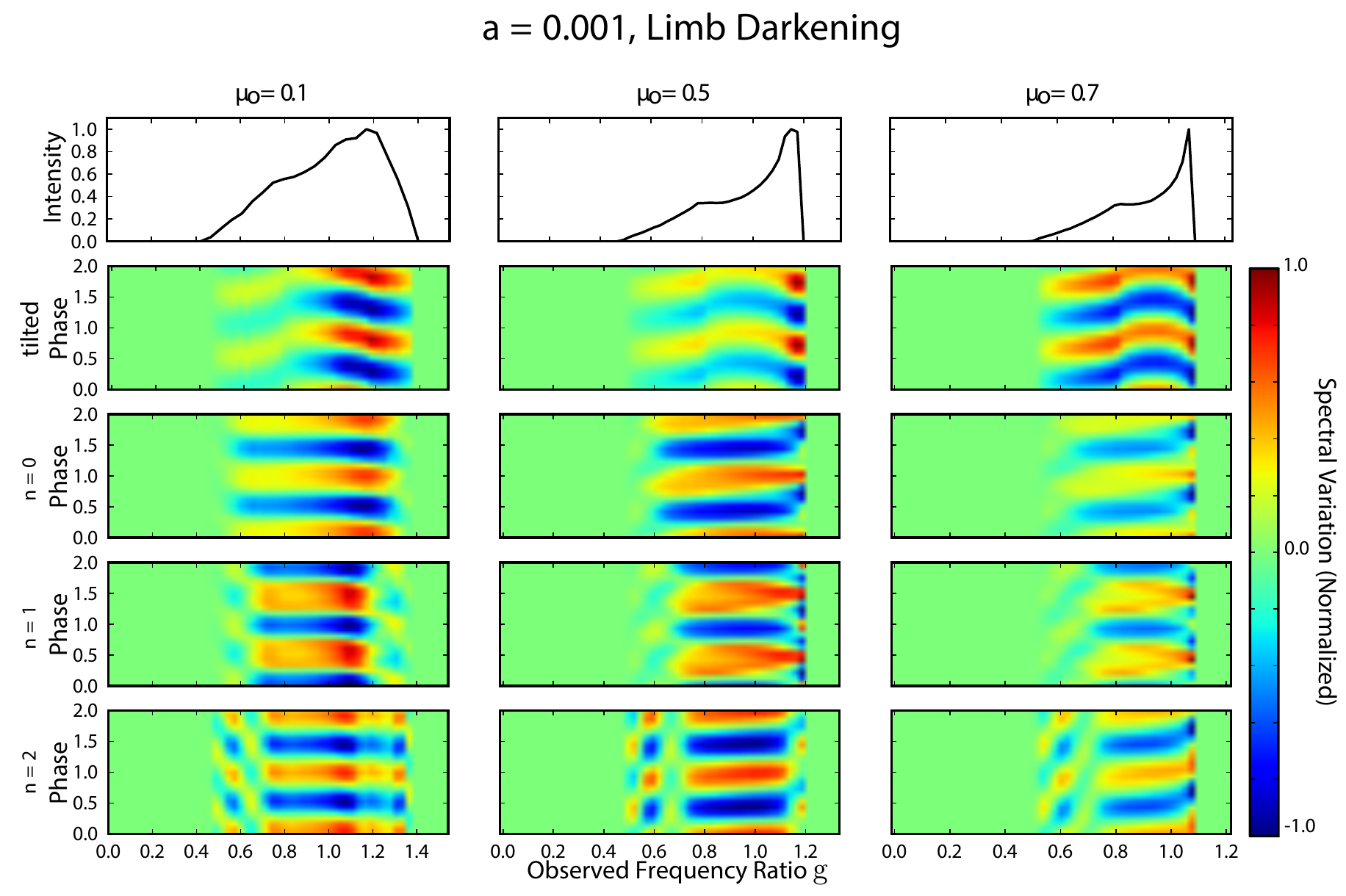}\\
\includegraphics[width=6.4in]{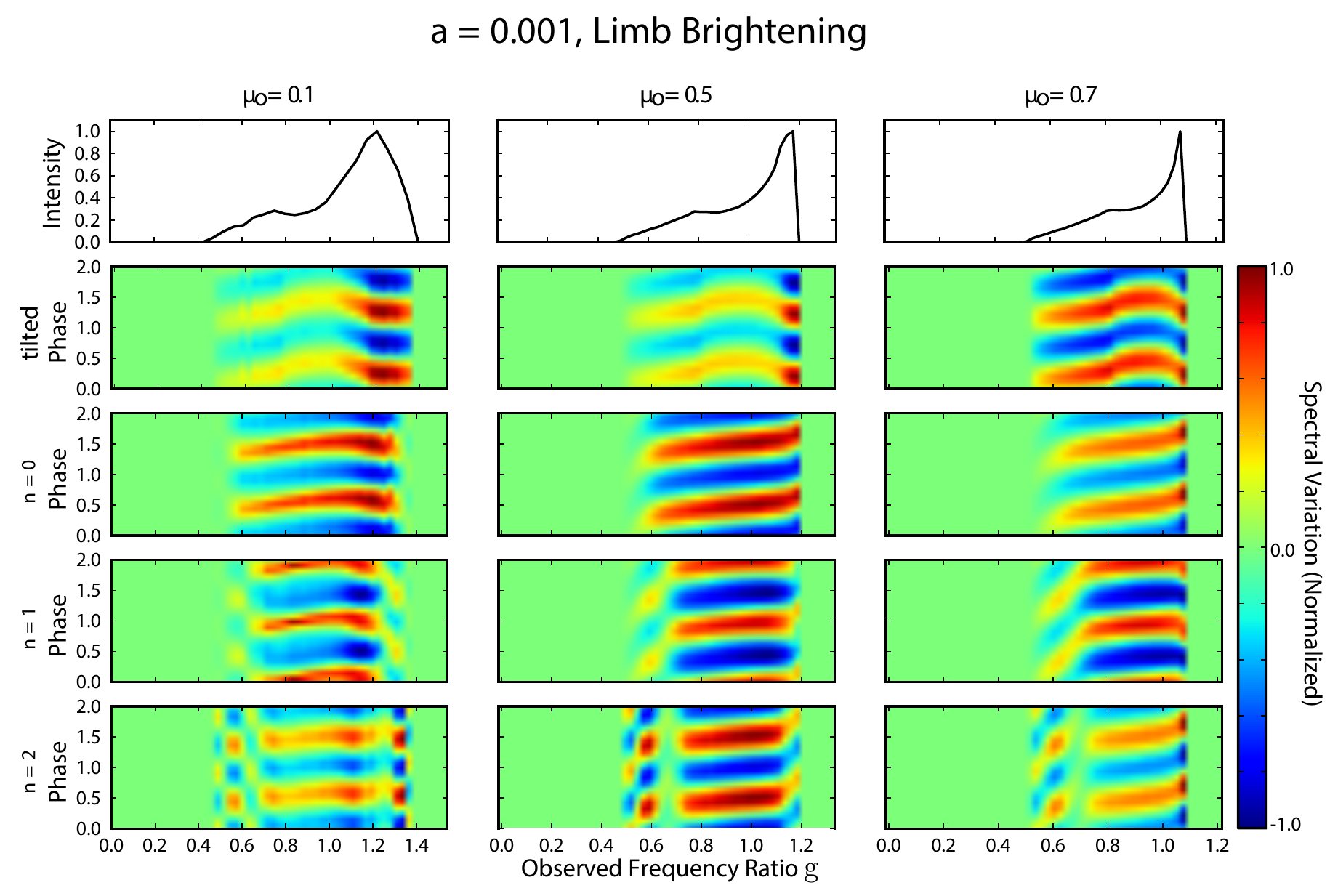}
\end{array}$
\end{center}
\caption{{\bf Upper panels}: Results assuming limb darkening angular emissivity for a black hole with spin $a = 0.001$ and inclinations $ \mu_{o} = \cos \theta_{\rm obs} = 0.1, 0.5$, and $0.7$. The first row shows the unperturbed broadened line spectra with the outer edge of the disc $r_{\rm out} = 20M$. The remaining rows show the spectral variation, the difference in intensity from the average spectra, as a function of oscillation phase and observed frequency ratio for the simple tilted disc, and the $n=0$, $n=1$, and $n=2$ corrugation modes. {\bf Lower panels:} Same as above, but for the limb brightening angular emissivity. \label{a1e-3.fig}}
\end{figure}

\begin{figure}
\begin{center}$
\begin{array}{c}
\includegraphics[width=6.2in]{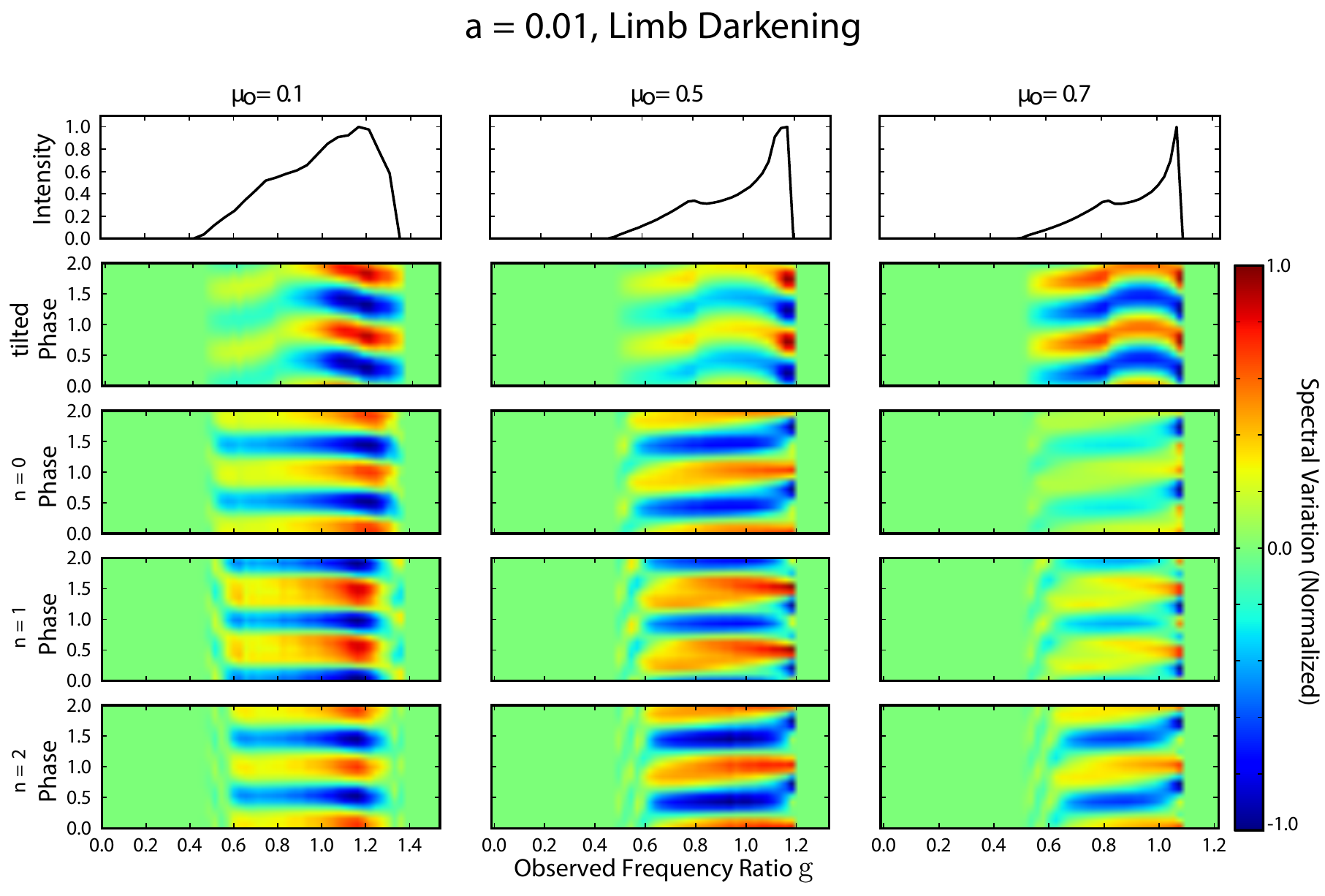}\\
\includegraphics[width=6.2in]{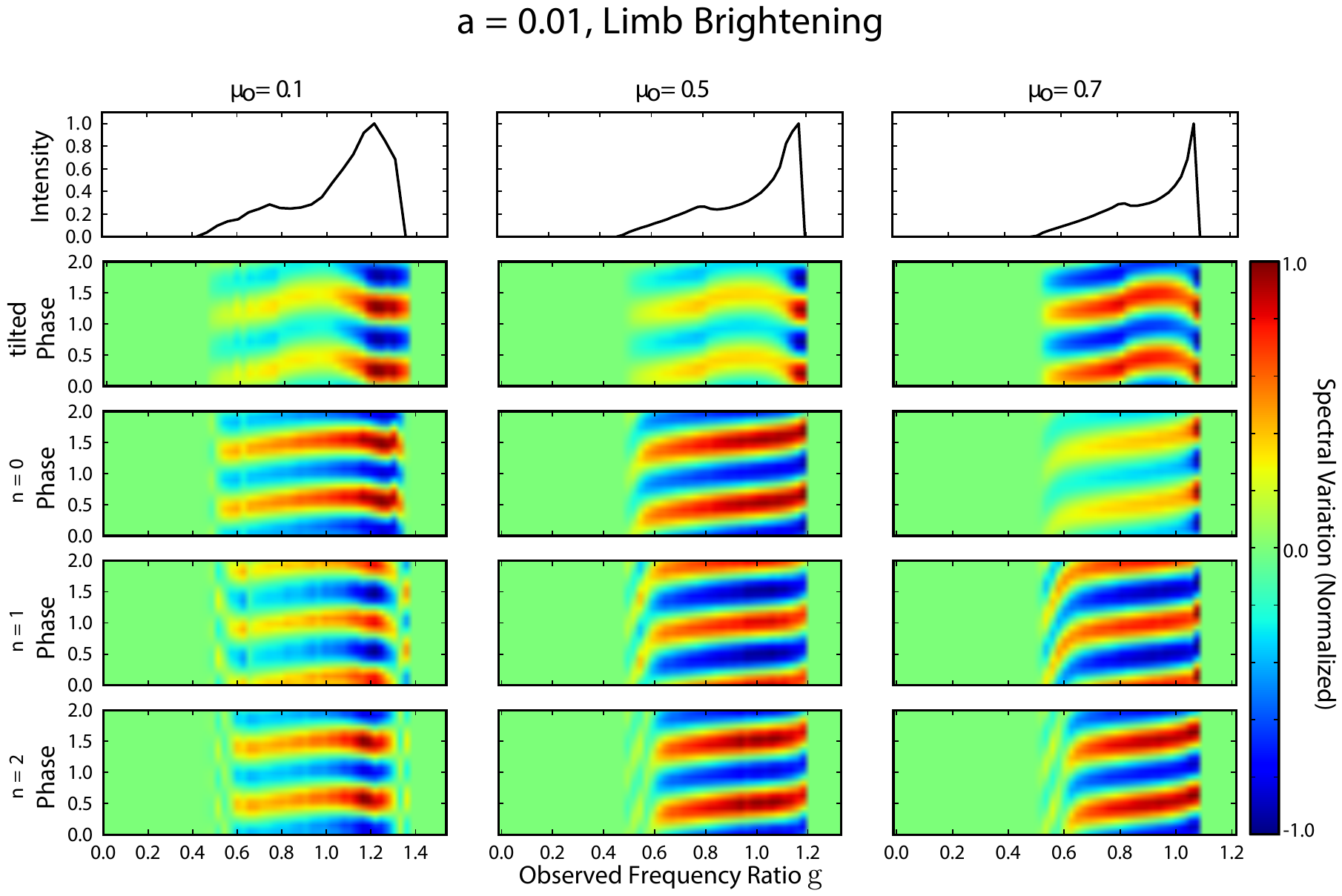}
\end{array}$
\end{center}
\caption{{\bf Upper panels}: Results assuming limb darkening angular emissivity for a black hole with spin $a = 0.01$ and inclinations $ \mu_{o} = \cos \theta_{\rm obs} = 0.1, 0.5$, and $0.7$. The first row shows the unperturbed broadened line spectra with the outer edge of the disc $r_{\rm out} = 20M$. The remaining rows show the spectral variation, the difference in intensity from the average spectra, as a function of oscillation phase and observed frequency ratio for the simple tilted disc, and the $n=0$, $n=1$, and $n=2$ corrugation modes. {\bf Lower panels:} Same as above, but for the limb brightening angular emissivity.\label{a1e-2.fig}}
\end{figure}

\begin{figure}
\begin{center}$
\begin{array}{c}
\includegraphics[width=6.2in]{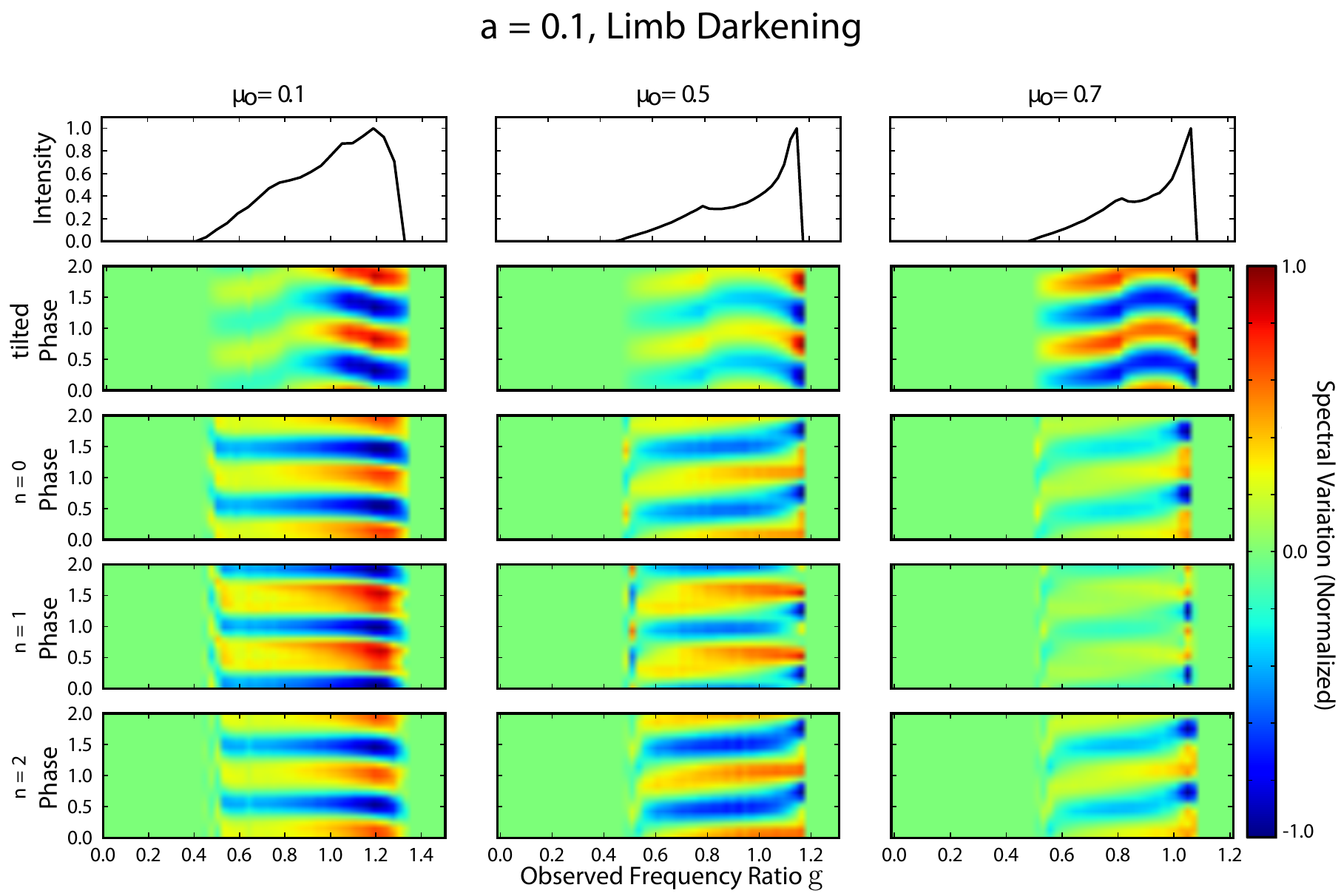}\\
\includegraphics[width=6.2in]{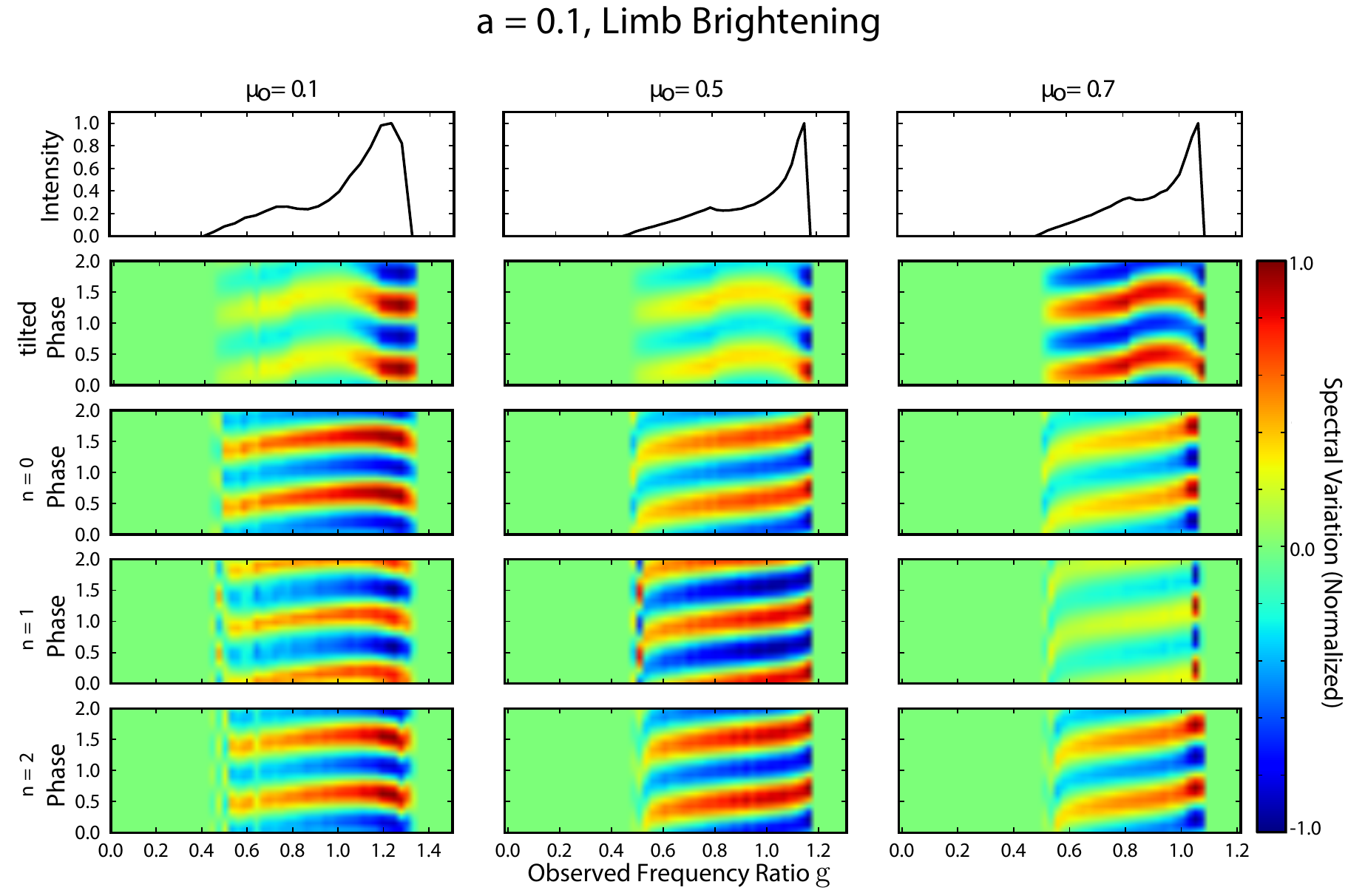}
\end{array}$
\end{center}
\caption{{\bf Upper panels}: Results assuming limb darkening angular emissivity for a black hole with spin $a = 0.1$ and inclinations $ \mu_{o} = \cos \theta_{\rm obs} = 0.1, 0.5$, and $0.7$. The first row shows the unperturbed broadened line spectra with the outer edge of the disc $r_{\rm out} = 20M$. The remaining rows show the spectral variation, the difference in intensity from the average spectra, as a function of oscillation phase and observed frequency ratio for the simple tilted disc, and the $n=0$, $n=1$, and $n=2$ corrugation modes. {\bf Lower panels:} Same as above, but for the limb brightening angular emissivity.\label{a1e-1.fig}}
\end{figure}

\begin{figure}
\begin{center}$
\begin{array}{c}
\includegraphics[width=6.2in]{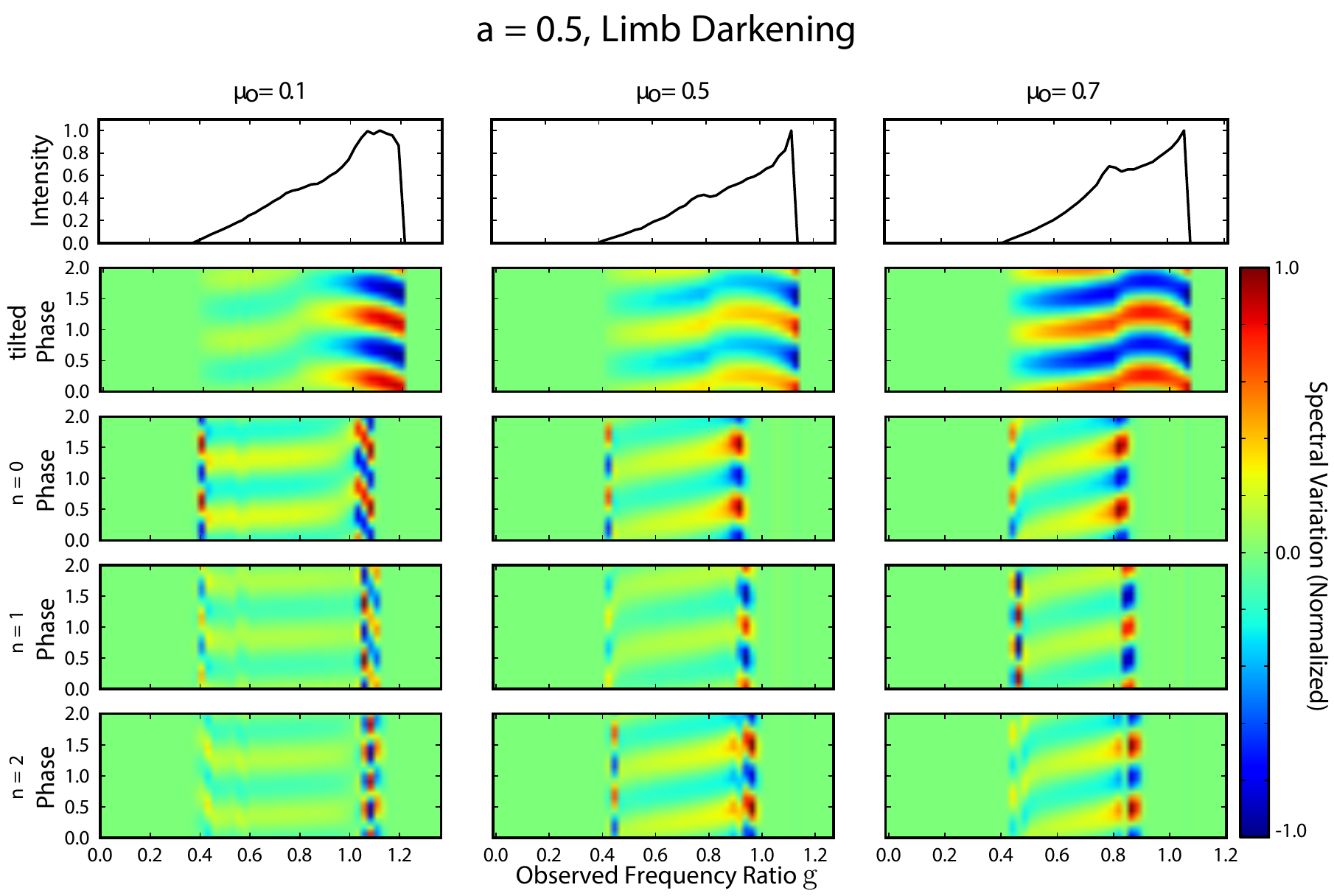}\\
\includegraphics[width=6.2in]{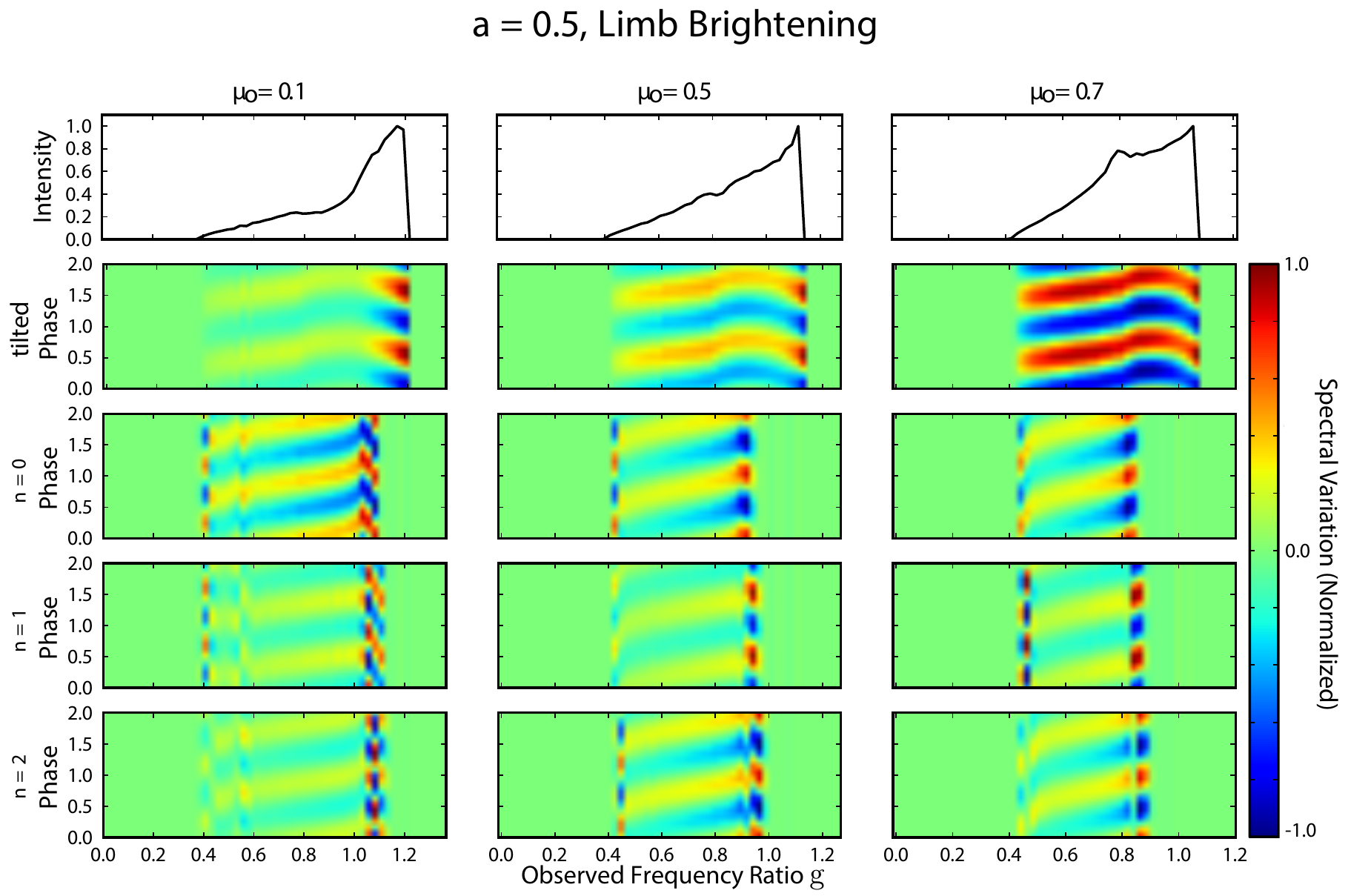}
\end{array}$
\end{center}
\caption{{\bf Upper panels}: Results assuming limb darkening angular emissivity for a black hole with spin $a = 0.5$ and inclinations $ \mu_{o} = \cos \theta_{\rm obs} = 0.1, 0.5$, and $0.7$. The first row shows the unperturbed broadened line spectra with the outer edge of the disc $r_{\rm out} = 20M$. The remaining rows show the spectral variation, the difference in intensity from the average spectra, as a function of oscillation phase and observed frequency ratio for the simple tilted disc, and the $n=0$, $n=1$, and $n=2$ corrugation modes. {\bf Lower panels:} Same as above, but for the limb brightening angular emissivity.\label{a5e-1.fig}}
\end{figure}

\begin{figure}
\begin{center}$
\begin{array}{c}
\includegraphics[width=6.2in]{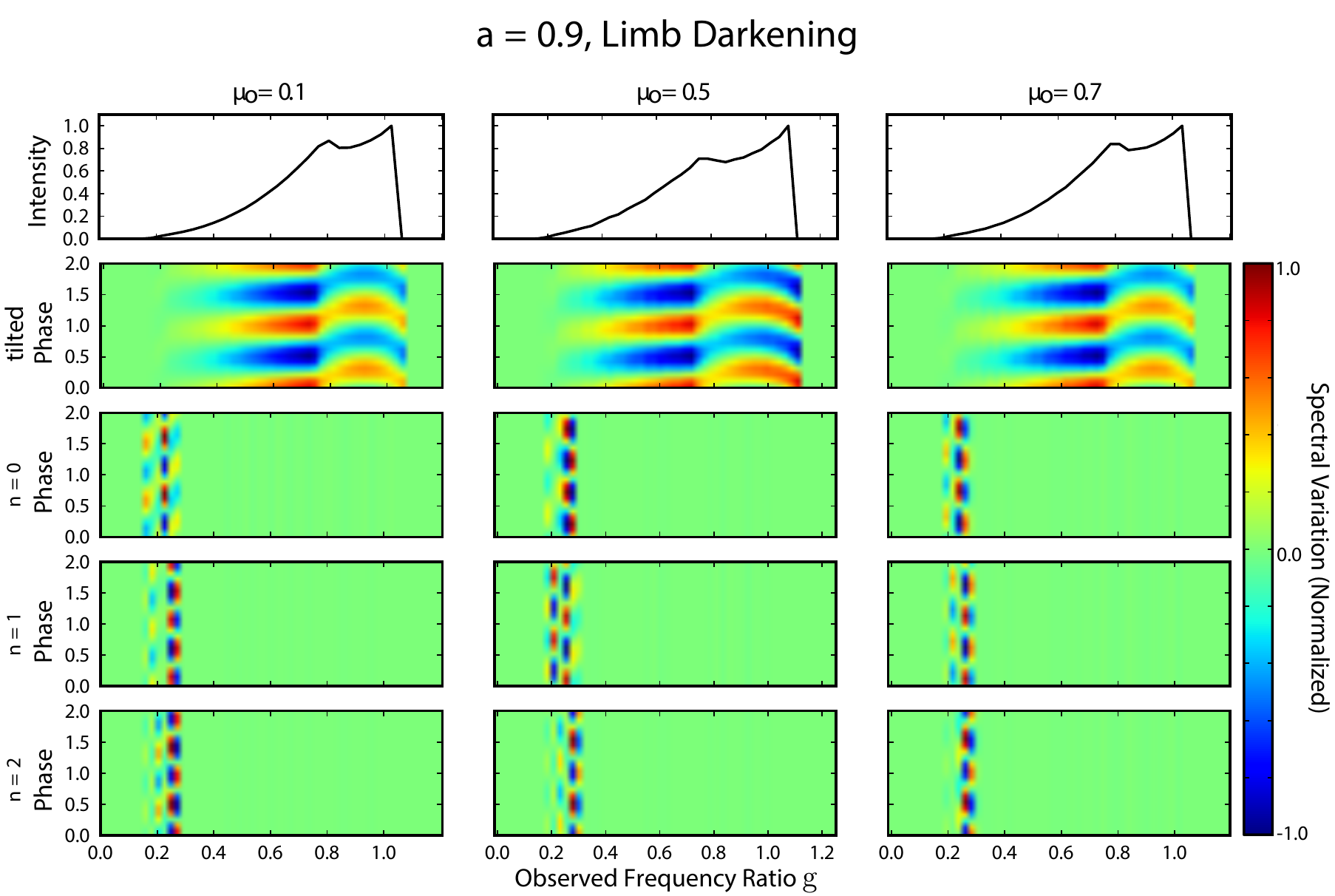}\\
\includegraphics[width=6.2in]{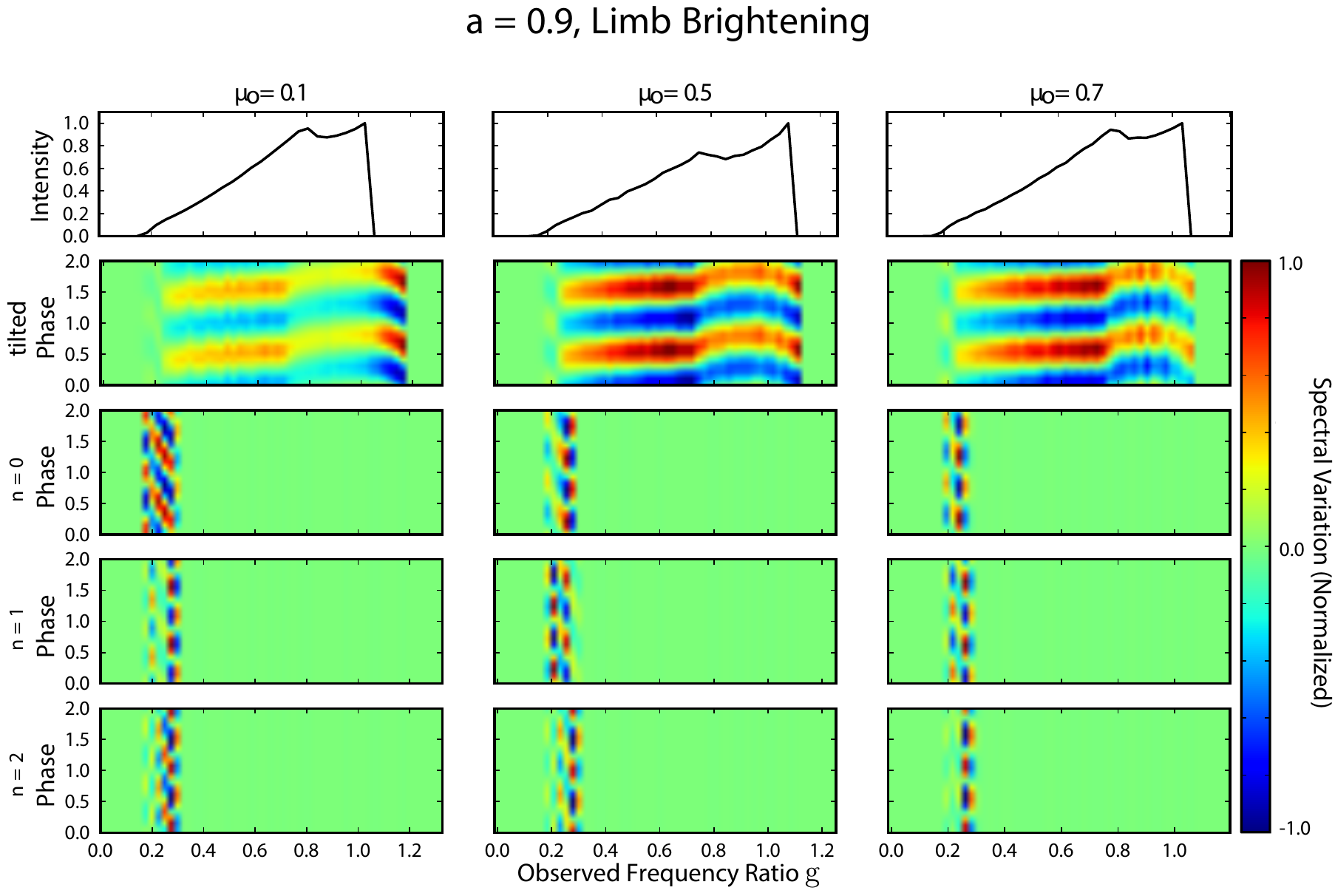}
\end{array}$
\end{center}
\caption{{\bf Upper panels}: Results assuming limb darkening angular emissivity for a black hole with spin $a = 0.9$ and inclinations $ \mu_{o} = \cos \theta_{\rm obs} = 0.1, 0.5$, and $0.7$. The first row shows the unperturbed broadened line spectra with the outer edge of the disc $r_{\rm out} = 20M$. The remaining rows show the spectral variation, the difference in intensity from the average spectra, as a function of oscillation phase and observed frequency ratio for the simple tilted disc, and the $n=0$, $n=1$, and $n=2$ corrugation modes. {\bf Lower panels:} Same as above, but for the limb brightening angular emissivity. The patchiness of the tilted-disc spectral variation in this case is due to mainly to resolution and redshift binning and is not physical. \label{a9e-1.fig}}
\end{figure}

For small amplitude perturbations spectral intensity plots were derived by binning the redshifts for each pixel, with number of bins dependent on the number of pixels subtended by the disc. Example spectra for various phases are shown in Figure \ref{specfig}, showing the spectral variability for different phases. This spectral variability is highlighted in Figures \ref{a1e-3.fig} - \ref{a9e-1.fig}. In these figures for a given black hole spin parameter $a$, observer inclination $\mu_o = \cos\theta_{\rm obs}$, and angular emissivity law we show the expected unperturbed spectrum of the broadened fluorescence line. We also calculate the normalized difference between the broadened line spectrum at a given oscillation phase and the phase-averaged mean spectrum, the spectral variation, and plot it as a function of oscillation phase and observed frequency ratio, $g$, for the simple tilted disc, and the $n=0, 1$ and $2$ c-modes. We normalize the spectral variation here since the amplitude of the perturbations is an arbitrary parameter in our formalism. 

\section{DIscussion}
Using a semi-analytic relativistic ray-tracing code for the Kerr metric we have calculated the time dependent line broadening signature of discoseismic corrugation modes in black hole accretion discs. We have shown that detailed spectral timing of the Fe-K$\alpha$ line variability could demonstrate if such corrugation modes are the source of LFQPOs observed in accreting black hole systems.

The spectral variability due to corrugation modes occurs in redshift ranges determined by the propagation region of the mode. In contrast the simple tilted-precessing disc model shows variability across the entire spectral range of the for the broadened line.

The redshift ranges where the variability of the corrugations modes manifests can be easily understood through the g-contours of Figures \ref{contour001} and \ref{contour5}. In Figure \ref{contour001} for $a = 0.001$ the propagation regions for the c-modes $r_{\rm ISCO} < r < r_{\rm IVR}$, span the range of constant g-contours. This can be directly compared to the spectral range of the variability in Figure \ref{a1e-3.fig}. For higher spins gravitational redshift of the line can begin to dominate the Doppler boosting due to Keplerian motion. The spectral range of the variability is reduced for $a = 0.5$, as shown in Figure \ref{a5e-1.fig}, and drastically reduced for the $a=0.9$ (Figure \ref{a9e-1.fig}) where the propagation region is much smaller and strongly dominated by gravitational redshift close to the black hole. The spectral range for variability in the $a = 0.5$ case can be compared directly to the g-contours spanned by the propagation region for the $n=2$ mode in Figure \ref{contour5}. 

While in principle the amplitude of the corrugation modes can be large, we have limited our detailed calculations to small amplitude perturbations as this simplifies the ray-tracing required to capture the effects on the Fe-K$\alpha$ emission. If amplitudes of the corrugation mode remain small, it would be extremely difficult to detect variability even with upcoming X-ray observatories.  However, the scaled spectral dependence of the variation as a function of phase should remain qualitatively similar for larger amplitude oscillations. Large amplitude corrugation modes can result in order unity variation of $\mu_{\rm em}$, or even self shadowing of the oscillation region, and are expected to lead to variations on the order of $\sim f{\rm var}$ of the flux in each spectral bin (see Figures \ref{fvar0.1.fig}-\ref{fvar0.7.fig}). While this varies significantly for different black hole spins and observer inclination, the reddest part of the iron line, where gravitational redshift dominates, will always emerge from the region closest to the black hole where $r_{\rm ISCO} < r < r_{\rm IVR}$. For bright sources, such as GRS $1915+105$, such large variations should be detectable given sufficient spectral ($\sim 0.05$ keV), and temporal resolution ($\sim 0.1$ s). For high-spin cases, the total fractional flux of the predicted variability decreases significantly, as the variability region shrinks, (see Figures \ref{contour001} and \ref{contour5}). Thus the corrugation mode variability of the iron line for high spin systems would be very difficult to detect. Additionally we note that corrugation modes in high spin systems ($a \go 0.5$) would have frequencies above most observed LFQPOs.

The signature of the radial order ($n$) of the oscillation is evident primarily in the red and blue edges of the spectral variation. The number of nodes in the oscillation mode is reflected in the number of nodes in the red or blue wings of the spectral variation at a particular phase, as seen in Figures \ref{a1e-3.fig} - \ref{a9e-1.fig}. This is particularly prominent for the low spin cases where the nodes are well spaced (Figures \ref{a1e-3.fig} - \ref{a1e-1.fig}). However, such fine signatures in the spectral variation would be difficult to detect unless sufficiently high spectral and timing resolution observations are stacked by oscillation phase. 

The presence or absence of such signatures in LFQPO-phase stacked Fe-K$\alpha$ observations can confirm or rule out corrugation modes as a source of LFQPOs. However, even if variability of the inner disc structure is not the source of the broad band LFQPOs, observations with high temporal and spectral resolution X-ray instruments, such as on the proposed LOFT or ATHENA missions, would allow iron line probes of inner disc structure variability. The spectral ranges and frequencies over which this variability is seen for corrugation modes can act as an probe of black hole spin parameter complementary to existing spin constraints from thermal and broadened iron-line observations. 

Our quantitative results have relied on raytracing for small amplitude oscillations. Detailed raytracing of large amplitude vertical displacements ($\xi_z \go H$) was beyond the scope of this paper, however the scaled spectral variability for large amplitudes should be qualitatively similar.  Finally we note that the disc models we have used have been for a simple barotropic thin disc, and should only be taken as a demonstration of physical principle. If such signatures are indeed observed, detailed modelling of the oscillation of more realistic discs and their vertical structure should be undertaken to constrain the parameters involved.

\section*{Acknowledgements}
DT was supported at Caltech by the Sherman Fairchild Foundation and at McGill through funding from the Canadian Institute for Advanced Research and the Lorne Trottier Chair in Astrophysics and Cosmology. IB conducted this research as part of a Summer Undergraduate Research Fellowship, and was supported by NASA ATP grant no. NNX11AC37G, NSF grant no. PHY-1151197, the David and Lucile Packard Foundation, the Alfred P. Sloan Foundation,  Mrs. Albert Burford, and the Sherman Fairchild Foundation. DT would like to acknowledge helpful discussion and useful advice from Sterl Phinney, Peter Goldreich, Christian Ott, Chris Hirata, Dong Lai, Marc Favata, Anil Zenigoulu, Chad Galley, and Andrew Cumming.

\appendix
\section{Ray-Tracing}\label{CodeAppendix}

In order to calculate various spectrum and light curve properties we must first construct a simulated image of the black hole and accretion disc in the observer's frame. In this frame the image is broken down into individual pixels of equal solid angle, and each corresponding to a single ray emitted by the accretion disc. At the observer each pixel can be indexed by the impact parameters $\alpha$($\perp$ to the spin axis projection), and $\beta$ ($\parallel$ to the spin axis projection).
disc
Solving for the geodesics, each of these rays can be back-traced to their source, allowing us to construct a complete image of the disc as seen by a distant observer. 

The contravariant components of photon momenta in a Kerr metric can be given in Boyer-Linquist coordinates, assuming $G=c=M_{BH}=1$ (e.g. Misner, Thorne \& Wheeler, 1973)
\ba
\biggl(\frac{dt}{d\lambda}\biggr) &= \rho^{-2}\biggl[\frac{r^2 + a^2}{\Delta}[E(r^2 + a^2) - L_z] - a(aE\sin^2\theta - L_z)\biggr],\\
\bigg(\frac{dr}{d\lambda}\biggr)&= \rho^{-2}[(E(r^2 + a^2)-L_za)^2 - \Delta((L_z - aE)^2 + Q)]^{1/2},\\
\biggl(\frac{d\theta}{d\lambda}\biggr) &= \rho^{-2}[Q - \cos^2\theta(L_z^2\csc^2\theta - E^2a^2)]^{1/2}, \\ 
\biggl(\frac{d\phi}{d\lambda}\biggr) &= \rho^{-2}\biggl[ -aE + L_z \csc^2\theta + \frac{a}{\Delta}(E(r^2+a^2) - L_za)\biggr],
\ea
where $a$ is the black hole spin, $\lambda$ is the affine parameter, and $\Delta = r^2 - 2r + a^2$. $E$, the photon energy, $L_z$ the angular momentum, and $Q$, Carter's constant, are constants of motion.

This form allows the use of simple Runge-Kutte routines to integrate out 
the photon paths, and are used by many authors as a compromise between code
complexity and computational speed. Care must be taken at the turning 
points of the $u$ and $\mu$ variables to ensure proper integration.

Utilizing elliptical integrals, and hence greater code complexity, Cunningham
and Bardeen (1973) outline a quicker method of calculating the photon 
trajectories using a Hamilton-Jacobi method. Here we follow a related procedure. Rearranging and switching variables from the affine parameter to a ``Mino parameter'' (see e.g. Drasco \& Hughes, 2004) $\lambda': \frac{d\lambda}{d\lambda'} = \rho^{-2}E$ we get the following coupled first order ODEs for the coordinates as a function of mino-parameter,
\ba
\biggl(\frac{dt}{d\lambda'}\biggr) &= T(r,\theta) \equiv \biggl[\frac{(r^2 + a^2)^2}{\Delta} - a^2\sin^2\theta \biggr] + a l \biggl[1 - \frac{r^2 + a^2}{\Delta}\biggr], \\
\bigg(\frac{dr}{d\lambda'}\biggr)^2 &= R(r) \equiv (r^2 + a^2 -la)^2 - (r^2 -2r + a^2)[(l-a)^2 +q^2], \\
\biggl(\frac{d\theta}{d\lambda'}\biggr)^2 &= \Theta(\theta) \equiv q^2 - l^2\cot^2\theta + a^2 \cos^2\theta,\\
\biggl(\frac{d\phi}{d\lambda'}\biggr) &= \Phi(r,\theta) \equiv l \csc^2\theta + a \biggl(\frac{r^2 + a^2}{\Delta} - 1 \biggr) - \frac{a^2 l}{\Delta}, 
\ea
where $l = L_z/E$, $q^2 = Q/E^2$. The constants of motion $l$ and $q^2$ are related to the impact parameters $(\alpha, \beta)$ by $l= -\alpha\sqrt{1-\mu_o^2}$ and $q^2 = \beta^2 + \mu_o^2(\alpha^2 - a^2)$. By utilizing the ``Mino parameter''  we avoid code complications involving integrating over multiple turning points in our dependent variable \citep[see e.g.][]{Dexter2009}. Our code performs similarly to that of \citet{Dexter2009}, which was used as a check for computational accuracy.

We can solve these ODE's in an semi-analytic fashion using the Jacobi elliptic functions and elliptic integrals. We first solve for the two independant variables, $r(\lambda')$ and $\theta(\lambda')$.

\subsection{The R equation}
We define the roots of the quartic equation
\be
R(r) = (r^2 + a^2 - al)^2 - (r^2 - 2r + a^2)[(l-a)^2 + q^2]=0,
\ee
$r_1, r_2, r_3$, and $r_4$ (see eg. Cadez et al., 2003), allowing us to rewrite the differential equation for $dr/d\lambda'$ as
\ba
\frac{dr}{d\lambda'} &= \sqrt{R(r)}\\
&= -\sqrt{(r-r_1)(r-r_2)(r-r_4)(r-r_3)}\\
\Delta \lambda' &= -\int_{r_o}^r \frac{dr}{\sqrt{(r-r_1)(r-r_2)(r-r_4)(r-r_3)}}\\
&=\frac{2}{\sqrt{(r_2-r_4)(r_1-r_3)}}\F\bigg(\sin^{-1}\sqrt{\frac{(r-r_1)(r_2-r_4)}{(r-r_2)(r_1-r_4)}}, \sqrt{\frac{(r_1-r_4)(r_2-r_3)}{(r_2-r_4)(r_1-r_3)}} \bigg)\bigg|_r^\infty\\
&= \frac{2}{\sqrt{(r_2-r_4)(r_1-r_3)}} \sn^{-1}\bigg( \sqrt{\frac{(r-r_1)(r_2-r_4)}{(r-r_2)(r_1-r_4)}}, \sqrt{\frac{(r_1-r_4)(r_2-r_3)}{(r_2-r_4)(r_1-r_3)}} \bigg) \bigg|_r^\infty,
\ea
where $\sn$ is the Jacobi elliptic function and $\F$ is the Jacobi elliptic integral of the first kind.\footnote{We calculate the Jacobi elliptic functions and elliptic integrals utilizing standard recurrence relations modified to work with values on part of the complex plane, combined with the Landen's transforms to change the complex arguments.}

With the negative value of the momentum corresponding to backtraced photon decreasing in $r$.
Solving for $r(\Delta \lambda')$ we get
\be
r(\Delta\lambda') = \frac{r_1(r_2-r_4)-r_2(r_1-r_4)\sn^2(u, \kappa_r)}{(r_2-r_4) - (r_1-r_4)\sn^2(u, \kappa_r)},
\ee
where
\ba
\kappa_r &= \sqrt{\frac{(r_1 - r_4)(r_2-r_3)}{(r_2-r_4)(r_1-r_3)}},\\
u &= \frac{\sqrt{r_2-r_4)(r_1-r_3)}\Delta\lambda'}{2} - u_\infty,\\
u_\infty &= \sn^{-1}\bigg(\sqrt{\frac{r_2 - r_4}{r_1 - r_4}}, \kappa_r \bigg),
\ea
even for complex values of $r_n$.

To calculate the value of $\Delta \lambda'(r)$ we must carefully consider any turning points that may be encountered in $r$. If any root $r_n$ is a positive real value greater than the horizon radius then the photon may have a turning point in $r$. If no turning point is encountered before the emission point then the value of $\Delta\lambda'(r)$ is given by
\be
\Delta\lambda'(r) = \frac{2}{\sqrt{(r_2-r_4)(r_1-r_3)}}\biggl[\F\bigg(\sin^{-1}\psi_\infty, \kappa_r\bigg)-\F\bigg(\sin^{-1}\psi(r), \kappa_r\bigg)\biggr],
\ee
where $\psi(r) = \sqrt{\frac{(r-r_1)(r_2-r_4)}{(r-r_2)(r_1-r_4)}}$ and $\psi_\infty = \psi(r)|_{r\rightarrow\infty} =\sqrt{\frac{(r_2-r_4)}{(r_1-r_4)}}$.

If a photon turning point is encountered by the backtrace before reaching the emission point (ie $\Delta \lambda' > \Delta\lambda'(r_{turn})$) then the corresponding value of $\Delta \lambda'$ is given by
\be
\Delta\lambda'(r) = \frac{2}{\sqrt{(r_2-r_4)(r_1-r_3)}}\bigg[\F\bigg(\sin^{-1}\psi(r), \kappa_r \bigg)\bigg|_{r_{turn}}^\infty + \F\bigg(\sin^{-1}\psi(r), \kappa_r \bigg)\bigg|_{r_{turn}}^{r_{\rm em}}\bigg].
\ee
As the change in sign corresponds to the change in the sign of the photon momenta at the turning point.

\subsection{The $\Theta$ equation}

Examining the $\theta$ equation we perform the substitution $z = \cos^2 \theta$ such that
\ba
\frac{d\theta}{d\lambda'} &= \pm \sqrt{\frac{-a^2z^2 - z[q^2 +l^2 -a^2] + q^2}{1- z}}\\
&= \pm \sqrt{\frac{a^2(z_+ - z)(z - z_-)}{1-z}},
\ea
where $z_\pm = -\frac{q^2 + l^2 -a^2}{2a^2} \pm \sqrt{\frac{(q^2 + l^2 -a^2)^2}{4a^4} + \frac{q^2}{a^2}}$
If we take $\chi: z = z_+ \cos^2\chi$ we have
\be
\frac{d\chi}{d\theta} = \pm \sqrt{\frac{1-z}{(z_+ - z)}},
\ee 
which gives
\ba
\frac{d\chi}{d\lambda'} = \frac{d\chi}{d\theta}\frac{d\theta}{d\lambda'} &= \sqrt{a^2(z - z_-)}\\
&= \sqrt{a^2(z_+ \cos^2\chi -z_-)}.
\ea
This gives the integral
\ba
 \lambda'(\chi) - \lambda'(\chi_o) &= \int_{\chi_o}^\chi \frac{d\chi}{\sqrt{a^2(z_+\cos^2\chi - z_-)}} \\
&= \frac{1}{a\sqrt{z+ - z_-}}\biggl[\F\biggl(\chi, \sqrt{\frac{z_+}{z_+-z_-}}\biggr)- \F\biggl(\chi_o, \sqrt{\frac{z_+}{z_+-z_-}}\biggr) \biggr],
\ea
where $\F(\psi, k)$ is the elliptic integral of the first kind. (In Abramowitz and Stegun notation this is $\F = \F(\psi|m)$, where $m = k^2$.)

Inverting this equation and solving for $\theta$ we finally obtain:
\be
\theta(\lambda') = \cos^{-1}(\sqrt{z_+} \cn(a\sqrt{z_+ - z_-}\lambda' + u_{\theta_o}, \kappa_\theta)),
\ee
where
\ba
u_{\theta_o} &= \sgn(\beta)\cn^{-1}(\cos(\theta_o)/\sqrt{z_+}, \kappa_\theta),\\
\kappa_\theta &= \sqrt{\frac{z_+}{z_+ - z_-}},
\ea
and $\cn(u, \kappa)$ is the Jacobi elliptic function which has inverse
$\cn^{-1}(u, \kappa) = \F(\cos^{-1}(u), \kappa)$. 

In order to calculate the mino-parameter corresponding to a particular value of $\theta$ we see
\be
\Delta\lambda'(\theta) = \frac{1}{a\sqrt{z_+ - z_-}}\biggl[\F\biggl(\chi, \sqrt{\frac{z_+}{z_+-z_-}}\biggr)-\F\biggl(\chi_o, \sqrt{\frac{z_+}{z_+-z_-}}\biggr)\biggr],
\ee
where 
\ba
\chi &= \cos^{-1}\biggl(\frac{\cos\theta}{\sqrt{z_+}}\biggr),\\
\chi_o &= \sgn(\beta)\cos^{-1}\biggl(\frac{\cos\theta_o}{\sqrt{z_+}}\biggr),
\ea
 thus we can solve for $\lambda'$ corresponding to intersection of the ray with simple fixed $\theta_{\rm em}$ disc configurations. The flat disc model of $\theta_{\rm em} = \pi/2$ is of particular interest.

\subsection{The $\Phi$ Equation}
The $\phi(\lambda')$ differential equation is more complicated than the equations for the first two spatial coordinates, however the differential equation can be solved by breaking the integration into two parts, integration over $\theta$ and integration over $r$. 

First rewriting the $\Phi(r, \theta)$ equation we see
\be
\frac{d\phi}{d\lambda'} = \frac{l}{1-\cos^2\theta} + a\frac{2r + la}{r^2 - 2r + a^2}.
\ee
Integrating the first term we see
\ba
\Delta\phi_1 &= \int_{\lambda'_o}^{\lambda_{\rm em}} \frac{l d\lambda'}{1 - \cos^2 \theta}\\
&= \int_{\lambda'_o}^{\lambda'_{\rm em}}\frac{l d\lambda'}{1 - z_+\cn^2(a\sqrt{z_+-z_-}\lambda' + u_{\theta_o}, k_\theta)}\\
&= \int_{\lambda'_o}^{\lambda_{\rm em}} \frac{ld\lambda'}{1 - z_+[1 - \sn^2(u, k_\theta)]}\\
&= \frac{l}{(1-z_+)}\int_{\lambda'_o}^{\lambda'_{\rm em}}\frac{d\lambda'}{1 + \frac{z_+}{1-z_+}\sn^2(u, k_\theta)}\\
&= \frac{l/a}{(1-z_+)\sqrt{z_+-z_-}}\int_{u_o}^{u_{\rm em}}\frac{du}{1 + \frac{z_+}{1-z_+}\sn^2(u, k_\theta)}\\
&= \frac{l/a}{(1-z_+)\sqrt{z_+-z_-}}\Pi\bigg(\chi, \frac{z_+}{1-z_+}, \kappa_\theta\bigg)\bigg|_{\chi_o}^{\chi_{\rm em}},
\ea
where $u = a\sqrt{z_+-z_-}\lambda' + u_o$, $\chi = \am(u, \kappa_\theta)$ is the Jacobi amplitude of $u$, and $\Pi(\psi, n, \kappa)$ is the elliptic integral of the third kind. All these values are real, and there is no difficulty in evaluating $\Pi(\psi, n, \kappa)$ through the standard recurrence relations.

The second term proves slightly more complicated:
\ba
\Delta\phi_2 &= a\int_{\lambda'_o}^{\lambda'_{\rm em}} \frac{(2r + la)d\lambda'}{r^2 - 2r + a^2}\\ 
&=a\int^{r_{\rm em}}_{\infty}\frac{2r + la}{(r-r_+)(r-r_-)}\frac{d\lambda'}{dr}dr\\
&= a\bigg(\frac{2r_+ + al}{r_+-r_-}\bigg)\int_{r_{\rm em}}^{\infty}\frac{1}{(r-r_+)}\frac{dr}{\sqrt{(r-r_1)(r-r_2)(r-r_3)(r-r_4)}}\nonumber\\
&\qquad \qquad- a\bigg(\frac{2r_- + al}{r_+-r_-}\bigg)\int_{r_{\rm em}}^{\infty}\frac{1}{(r-r_-)}\frac{dr}{\sqrt{(r-r_1)(r-r_2)(r-r_3)(r-r_4)}}\\
&= a\bigg[\frac{2r_2 + al}{r_2^2-2r_2 + a^2}\bigg]\Delta\lambda'+ \frac{2a}{\sqrt{(r_1-r_3)(r_2-r_4)}}\bigg(\frac{r_1-r_2}{r_+-r_-}\bigg)\nonumber\\
&\qquad \qquad \qquad\times \bigg[\frac{2r_-+al}{(r_1-r_-)(r_2-r_-)}\Pi_- - \frac{2r_++al}{(r_1-r_+)(r_2-r_+)}\Pi_+\bigg]_{r_{\rm em}}^{\infty}
\ea
where 
\ba
r_\pm &= 1 \pm \sqrt{1-a^2}\\
\Pi_\pm &= \Pi(\psi(r), n_\pm, \kappa_r)\\
n_\pm &= \psi^2(r)\bigg|_{r=r_\pm}.
\ea

Thus we have 
\be
\Delta\phi(\Delta\lambda') = \Delta\phi_1 + \Delta\phi_2.
\ee
Note that when a turning point is encountered in r, the $\Delta\phi_2$ must be evaluated with the appropriate sign change as for $\Delta\lambda'(r)$.

In general the parameters of the elliptical integrals of the third kind will be complex. This limits evaluation of $\Delta\phi_2$ using recurrence relations to a particular domain of complex parameter space. Evaluations outside of this domain can be performed using numerical quadrature.

\subsection{The T Equation}\label{TeqAppendix}
The solution to the time coordinate is the most complicated of the four Boyer-Linquist coordinates. Simplifying the $T(r, \theta)$ equation we have:
\be
\frac{dt}{d\lambda'} = - a^2\sin^2\theta + \frac{(r^2 + a^2)^2 + 2alr}{r^2 - 2r + a^2}.
\ee
The first term can be integrated in a relatively straightforward fashion as:

\ba
\Delta t_1 &= \int a^2(1 + \cos^2\theta) d\lambda'\\
&= a^2\Delta\lambda' - \int a^2 z_+ \cn^2(u_\theta, \kappa_\theta))\frac{du_\theta}{a\sqrt{z_+-z_-}}\\
&= a^2 \Delta\lambda' - \frac{az_+}{\sqrt{z_+-z_-}}\frac{1}{\kappa_\theta^2}\bigg[\E(\am(u_\theta), \kappa_\theta) - (1-\kappa_\theta^2)u_\theta\bigg]_{u_o}^{u_{\rm em}}\\
&=a^2\bigg[a + \frac{z_+(1-\kappa_\theta^2)}{\kappa_\theta^2}\bigg]\Delta\lambda' - \frac{az_+}{\kappa_\theta^2\sqrt{z_+-z_-}}\E(\chi, \kappa_\theta)\bigg|_{\chi_o}^{\chi_{\rm em}}.
\ea
For the second term,$\Delta t_2\equiv \frac{(r^2 + a^2)^2 + 2alr}{r^2 - 2r + a^2}$,  the analytic solution can be determined as a very long combination of elliptical integrals of the first, second and third kinds as well as the Jacobi elliptic functions. 

However, the values of observer time elapsed for the intervals between photon emission and observation at infinity are necessarily divergent. Though one could simply evaluate $\Delta t_2$ only up to a large arbitrary value of $r$, it is better to instead subtract off the same infinite constant for each ray, as our interest is limited to the elapsed observer time difference between different rays, by considering the Kerr time.

Remembering that the time in Kerr-coordinates is given by a transformation:
\be
dt_{\rm Kerr} = dt_{\rm BL} + \bigg(\frac{r^2 + a^2}{r^2-2r+a^2}\bigg) dr
\ee
we can then subtract off the same constant for each ray by taking
\ba
\Delta t' &= \Delta t_{\rm Kerr} - \int_{r_{\rm em}}^{r_{\rm bitrary}}\frac{r^2 + a^2}{r^2-2r + a^2} dr\\
&= \Delta t_{\rm BL} + \int_\infty^{r_{\rm em}} \frac{r^2 + a^2}{r^2 - 2r + a^2}dr  - \int_{r_{\rm em}}^{r_{\rm bitrary}}\frac{r^2 + a^2}{r^2-2r + a^2} dr\\
&= \Delta t_1 + \Delta t_2 - \int_{r_{\rm bitrary}}^\infty\frac{r^2 + a^2}{r^2 -2r + a^2}dr \\
&= \Delta t_1 + \int_{r_{\rm em}}^{r_{\rm bitrary}} \frac{(r^2 + a^2)^2 + 2alr}{r^2 - 2r + a^2}\frac{dr}{\sqrt{R(r)}} \nonumber\\
&\qquad \qquad \qquad + \int_{r_{\rm bitrary}}^\infty \bigg[\frac{(r^2 + a^2)^2 + 2alr}{r^2 - 2r + a^2}\frac{1}{\sqrt{R(r)}} - \frac{r^2 + a^2}{r^2 - 2r + a^2}\bigg]dr, \label{6.70}
\ea
where $r_{\rm bitrary}$ is some $r$ beyond the disc range. 

These integrations are best performed with numerical quadrature as evaluation of the elliptic integrals and Jacobi elliptic functions prove less efficient than the numerical integration. In addition the $1/\sqrt{R(r)}$ term makes it difficult to remove the divergent components of the analytic solution so it may be evaluated numerically.

\renewcommand{\bibsection}{\section*{References}} 

\end{document}